\documentclass[longbibliography, aps, prx, amsmath, amssymb, amsfonts, twocolumn,superscriptaddress, footinbib]{revtex4-1}
\usepackage{mathrsfs}
\usepackage{amssymb}
\usepackage{graphicx}
\usepackage{dcolumn}
\usepackage{bm}
\usepackage{bbold}
\usepackage{textcomp}
\usepackage{amsmath}
\usepackage[titletoc]{appendix}
\usepackage{tabularx}
\usepackage{verbatim}
\usepackage{hyperref}
\hypersetup{                    
	colorlinks,
	citecolor=blue,
	filecolor=blue,
	linkcolor=blue,
	urlcolor=blue
}

\newtheorem{remark}{Remark}
\newtheorem{Def}{Definition}

\newcommand\ket[1]{\ensuremath{|#1\rangle}}

\newcounter{RomanNumber}

\begin{document}
	\title{Quantum Digital Signatures with Random Pairing}
	
	\author{Ji-Qian Qin}
	\affiliation{ State Key Laboratory of Low Dimensional Quantum Physics, Department of Physics, \\
	Tsinghua University, Beijing 100084, China}
	\author{Cong Jiang}
	\affiliation{ State Key Laboratory of Low Dimensional Quantum Physics, Department of Physics, \\ Tsinghua University, Beijing 100084, China}
	\affiliation{ Jinan Institute of Quantum technology, SAICT, Jinan 250101, China}
	\author{Yun-Long Yu}
	\affiliation{ State Key Laboratory of Low Dimensional Quantum Physics, Department of Physics, \\ Tsinghua University, Beijing 100084, China}
	\author{Xiang-Bin Wang}
	\email{ xbwang@mail.tsinghua.edu.cn}
	\affiliation{ State Key Laboratory of Low Dimensional Quantum Physics, Department of Physics, \\ Tsinghua University, Beijing 100084, China}
	\affiliation{ Jinan Institute of Quantum technology, SAICT, Jinan 250101, China}
	\affiliation{ Shanghai Branch, CAS Center for Excellence and Synergetic Innovation Center in Quantum Information and Quantum Physics, University of Science and Technology of China, Shanghai 201315, China}
	\affiliation{ Shenzhen Institute for Quantum Science and Engineering, and Physics Department, Southern University of Science and Technology, Shenzhen 518055, China}
	\affiliation{ Frontier Science Center for Quantum Information, Beijing, China}

\begin{abstract}
\noindent	Digital signatures can guarantee the unforgeability and transferability of the message. Different from classical digital signatures, whose security depends on computational complexity, quantum digital signatures (QDS) can provide information-theoretic security. We propose a general method of random pairing QDS (RP-QDS), which can drastically improve QDS efficiency. In a way, our random pairing method provide a tightened result of security level of QDS. In the method, the parity value of each pair is used for the outcome bit value. We present general formulas for fraction of untagged bits and error rates of the outcome bits. Random pairing can be applied as a fundamental method to improve the QDS efficiency for all existing quantum key distribution (QKD) protocols. We take sending-or-not-sending (SNS) QDS and side-channel-free (SCF) QDS as examples to demonstrate the advantage of random pairing through numerical simulation. Similar advantage with random pairing is also founded with decoy-state MDIQKD and also decoy-state BB84 protocol. We study the RP-SNS-QDS with finite data size through novel optimization. The numerical simulation results show that the signature rate can be increased by more than $100\%$ under noisy channel using our random pairing method. 
\end{abstract}

\maketitle
\section{Introduction}

Digital signatures can guarantee classical messages to be securely exchanged from one signer to multiple receivers~\cite{1055638}. 
Classical digital signatures include signature and verification algorithms and the security is based on computational complexity. In contrast,
quantum digital signatures (QDS)~\cite{gottesman2001quantum} can provide a higher level of security, information-theoretic security, which is guaranteed by the fundamental principles of quantum mechanics. Potentially, QDS is practically useful because the original assumptions of quantum memory and secure quantum channel have been removed~\cite{AnderssonQDS2006,dunjko2014quantum,amiri2016secure,yin2016practical}.  Actually, a lot of experimental demonstrations have been done for  QDS ~\cite{collins2014realization,donaldson2016experimental,croal2016free,yin2017experimental,roberts2017experimental,zhang2018proof,ding2020280} in applications such as emails, financial transactions, and more~\cite{pirandola2020advances}.

In this work, we mainly consider the protocol with one signer (Alice) and two receivers (Bob and Charlie). It should be noted that, in the three-party protocol, at most one party can be dishonest. Because if there are two dishonest parties, it is easy to cheat on the third party through collusion. The security in QDS means that the message has unforgeability and transferability~\cite{gottesman2001quantum}. A successful forgery means that anyone other than Alice can produce a signature which is accepted by receivers. Transferability (or nonrepudiation) means that if one honest receiver accepts the signature generated by Alice, the other honest receiver will also accept this signature. 
The QDS protocols~\cite{wallden2015quantum,amiri2016secure,puthoor2016measurement} usually have two stages: the first is the distribution stage, where Alice-Bob and Alice-Charlie independently perform key generation protocol (KGP) to generate the correlated bit strings for signature; the second is the messaging stage, where the message is signed and verified.

Given the prior art theory~\cite{amiri2016secure}, the quantum communication of any quantum key distribution (QKD) protocol can be directly applied for QDS through taking the post data processing on the raw bits there. Given the quantum part of various QKD protocols, for example, the BB84~\cite{BENNETT20147}, the decoy-state BB84~\cite{Hwang2003,Wangdecoy2005,Lodecoy2005,PhysRevLett.121.190502}, the measurement-device-independent (MDI) QKD~\cite{lo2012measurement,BraunsteinMDI2012,WangThreeMDI2013,zhouFourMDI2016,MDI404}, sending-or-not-sending (SNS)~\cite{wang2018twin} of twin-field (TF) QKD~\cite{LuTF2018} and so on, we can apply the existing post data processing method~\cite{amiri2016secure} for QDS. Yet, it is still an interesting problem to find novel post data processing method to improve the QDS efficiency. Here in this work, we propose such a method, the novel method of random pairing QDS (RP-QDS). The method can be applied to all existing QKD protocols and improve the QDS efficiency drastically. In real situations, the data size is always finite. We consider RP-SNS-QDS with finite pluses and optimize the signature rate with finite data size.

\section{Iteration relation of random pairing} \label{s2}
Compared with other QDS protocols, the biggest difference of our protocol is that the step of random pairing is applied in the KGP, which can effectively improve the signature efficiency. 

\begin{figure}
	\centering
	\includegraphics[scale=0.5]{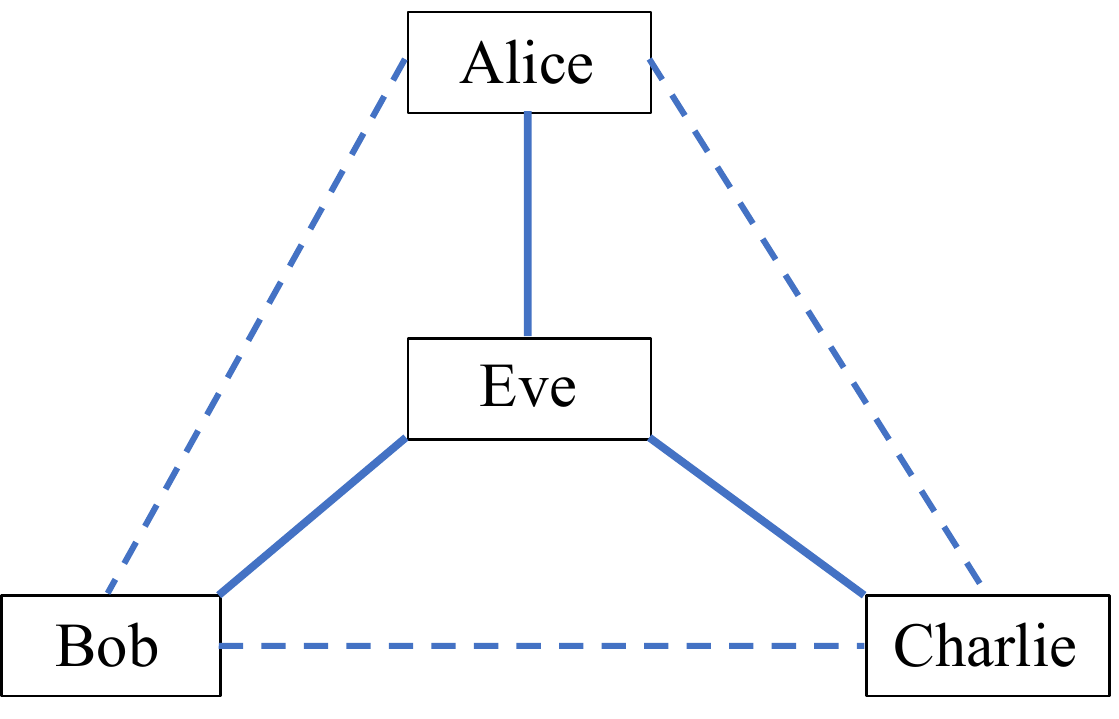}
	\caption{Schematic of quantum digital signature~\cite{puthoor2016measurement,yin2017experimental,WangTFQDS2021}. Our random pairing method can be directly used in the existing quantum digital signature protocols. Alice can sign the message and send it with the signature to Bob and Charlie for verification. Eve is a relay and can be un-trusted. Solid lines and dashed lines represent quantum channels and authenticated classical channels, respectively.}
	\label{QDS}
\end{figure}

As shown in Fig.~\ref{QDS}, there are quantum channels (solid lines) between each party (Alice, Bob and Charlie) and the relay Eve, and there are authenticated classical channels (dashed lines) between each two of parties. It should be noted that in the KGP, depending on the needs of specific protocols, such as MDI-QDS~\cite{puthoor2016measurement}, a relay Eve can be added, who can be un-trusted. In our protocol, Alice is a signer and Bob and Charlie are two receivers. 
Alice can sign the message and send it with signature to two receivers for verification. 

We need to determine the bit-flip error rate and phase phase-flip error rate of the bit string after random pairing to choose the length of the signature string, which can affect the security level and efficiency of the protocol.

In the KGP of distribution stage after error test, Alice and Bob obtain the $N_t$-bit string $z_A$ and $z_B$ with bit-flip error rate $E$ (error rate in $Z$ basis) and phase-flip error rate $e^{\mathrm{ph}}$ (error rate in $X$ basis). They take the random pairing and use the parity values of the bit pairs as the outcome of each pairs. The outcome bits make new bit strings $z_A^{\prime}$ and $z_B^{\prime}$ at each sides with each of them containing $N_t/2$ bits. They use the new bit
strings $z_A^{\prime}$ and $z_B^{\prime}$ for signature. We can list the following 
steps for the method: 1, After KGP with error test, they obtain $N_t$ raw bits. Bob randomly pairs his bits and asks Alice to take the same pairing. 2, They use the parity value of each pairs as the outcome bits. For example, suppose a pair containing bit $i$ and bit $j$ in the initial string $z_A$ ($z_B$)  with bit values $z_{A,i}$ and $z_{A,j}$ ($z_{B,i}$ and $z_{B,j}$). Bob (Alice) uses parity value $z_{B,i}\oplus z_{B,j}$ ($z_{A,i}\oplus z_{A,j}$) for the outcome bit value, where notation $\oplus$ denotes bit addition modulo $2$ and $i,j\in\{1,2,\dots,N_t\}$, $i\neq j$. After this step they obtain $N_t/2$ outcome bits. 3, They calculate the phase error rate of those $N_t/2$ outcome bits, test their bit-flip error rate and use them for QDS.

After random pairing, the number of effective bits $N_t^{\prime}$ becomes half of the original 
\begin{equation}\label{Nt}
	\begin{split}
		&N_{t}^{\prime}=\frac{1}{2}N_{t},\\
	\end{split}
\end{equation}
where $N_t$ represents the number of effective bits before random pairing. Without loss of generality, we always assume $N_t$ to be an even number.

The bit-flip error rate $E^{\prime}$ for the $N_t^{\prime}$ outcome bit string after random pairing can be tested directly. When only one of the paired bits in original bit string has an error, the new bit made up of their parity value will have an error. So the expected value for bit flip error rate $E^{\prime}$ of new bit strings is

\begin{equation}\label{Ep}
	\begin{split}
		&E^{\prime } = 2E(1-E),\\
	\end{split}
\end{equation}
where $E$ represents the bit flip error rate before random pairing. Note that the bit-flip error of the outcome bits after random pairing is supposed to be tested directly in our protocol as shown later.  Eq.~\eqref{Ep} above is the expected[tation value] result used for numerical simulation.

The untagged bit in new bit string is generated by pairing two untagged bits or one untagged bit with one tagged bit in original bit string, so the proportion $\Delta_{\mathrm{un}}^{\prime}$ of untagged bits (which are  defined in specific QDS protocols in next section) after random pairing is
\begin{equation}
	\begin{split}
		&\Delta_{\mathrm{un}}^{\prime}=\Delta_{\mathrm{un}}^2+2\Delta_{\mathrm{un}}(1-\Delta_{\mathrm{un}}),\\
	\end{split}
\end{equation}	
where $\Delta_{\mathrm{un}}$ represents the proportion of untagged bits before random pairing.

\begin{figure}
	\centering
	\includegraphics[scale=0.5]{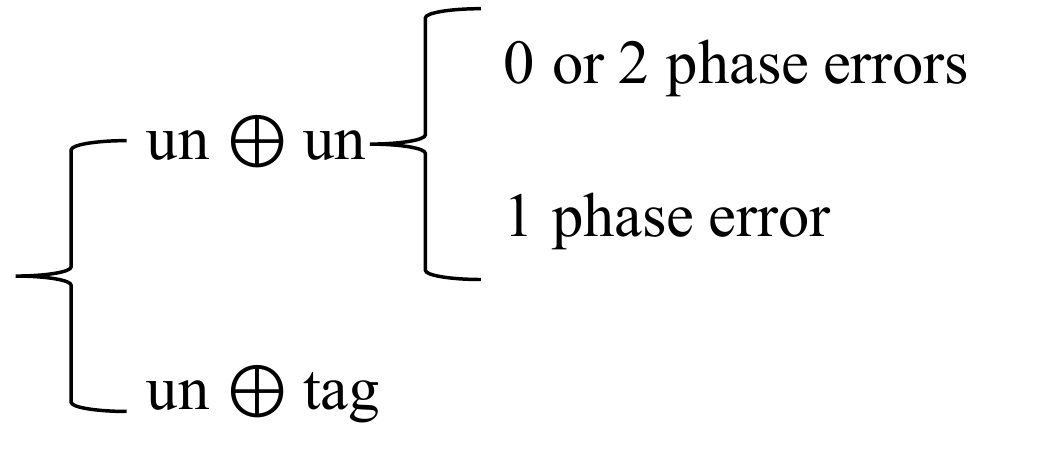}
	\caption{The phase flip error rate after random pairing. Notations "un $\oplus$ un" and "un $\oplus$ tag" represent two untagged bits are paired and one untagged bit and one tagged bit are paired, respectively.}
	\label{rp}
\end{figure}

As shown in Fig.~\ref{rp}, we analyze the phase flip error rate after the random pairing in two cases: one is two untagged bits are paired with probability $\Delta_{\mathrm{un}}^2$, the other is one untagged bit and one tagged bit are paired with probability $2\Delta_{\mathrm{un}}(1-\Delta_{\mathrm{un}})$. 

\begin{figure}
	\centering
	\includegraphics[scale=0.35]{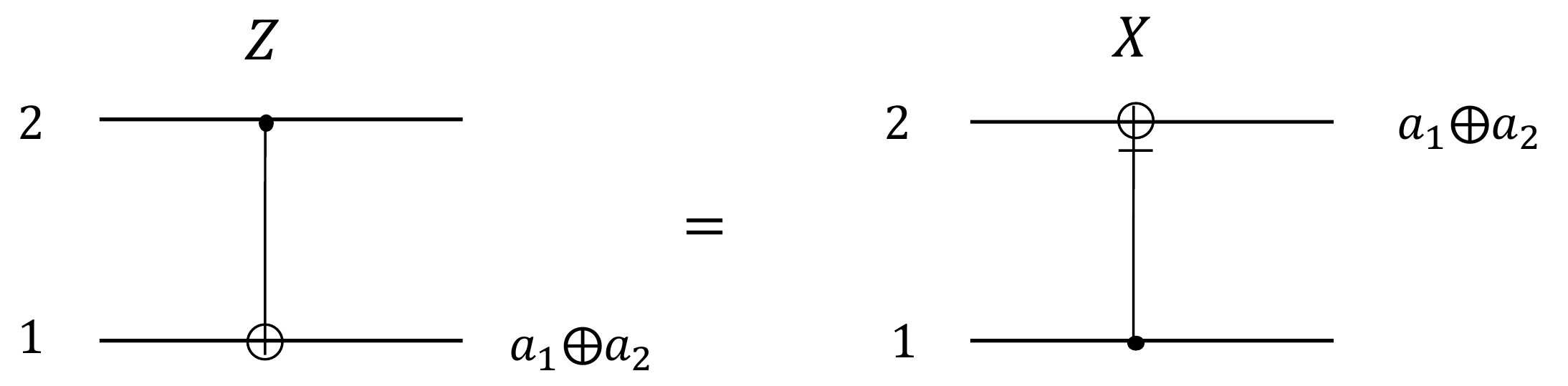}
	\caption{The CNOT operation in $Z$ basis is equivalent to the inverse CNOT operation in $X$ basis \cite{GottesmanLo2002TWCCQKD}. }
	\label{p_eph}
\end{figure}

\begin{figure}
	\centering
	\includegraphics[scale=0.35]{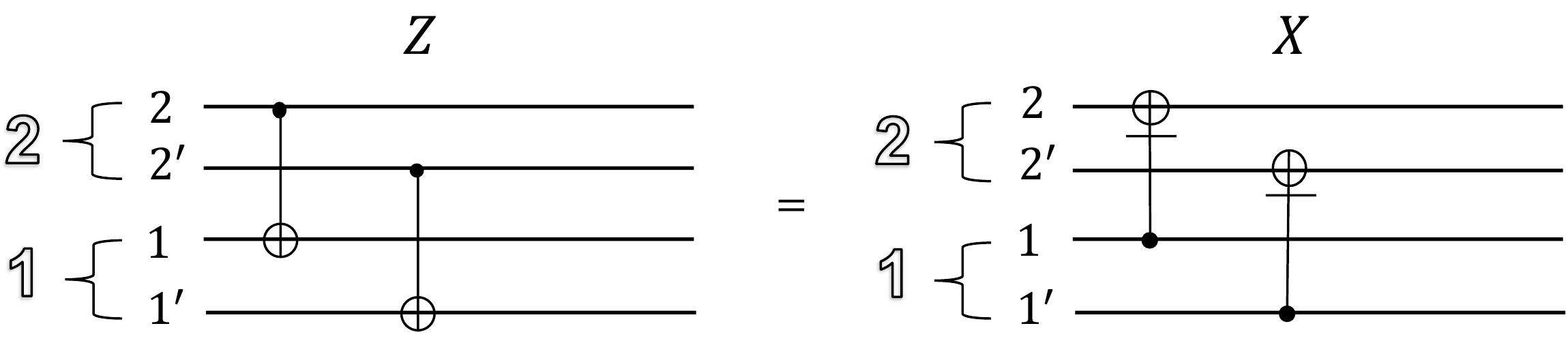}
	\caption{The bilateral CNOT operations on two entanglement pairs, entanglement pair $\mathbb{1}$ and entanglement pair $\mathbb{2}$, in the virtual protocol. The qubit $1$ and $2$ belong to Alice and the qubit $1^{\prime}$ and $2^{\prime}$ belong to Bob.}
	\label{p_eph2}
\end{figure}

When one untagged bit and one tagged bit are paired, the phase flip error rate is
\begin{equation}
	\tilde{e}^{ \mathrm{ph}}=e^{\mathrm{ph}}.
\end{equation}

For the iteration formulas of phase error for the outcome bits from pairs containing two untagged bits, we consider the virtual protocol using quantum entanglement. In the real protocol, there is a pair containing two untagged bits, bit $1$ with bit value $a_1$ and bit $2$ with bit value $a_2$. As shown in Fig.~\ref{p_eph}, the classical operation of $a_1 \oplus a_2$ in the real protocol is equivalent to the CNOT operation in $Z$ basis on the two qubits, qubit $1$ and qubit $2$ in the pair. This is also equivalent to the CNOT operation in $X$ basis with the target bit and control bit reversed \cite{GottesmanLo2002TWCCQKD}. After this operation, qubit $2$ is discarded and qubit $1$ is used for further processing.

The CNOT operation above corresponds to bilateral CNOT operations on two entanglement pairs in the virtual protocol as shown in Fig.~\ref{p_eph2}. Consider the corresponding virtual protocol using quantum entanglement in Fig.~\ref{p_eph2}. After the bilateral CNOT operations, the phase error of the entanglement pair $\mathbb{2}$ is determined by the number of initial phase errors of those two entanglement pairs, i.e., $0$ or $2$ phase errors will not cause any phase error in entanglement pair $\mathbb{2}$, while $1$ phase error will cause one error of the entanglement pair $\mathbb{2}$. This fact means after the bilateral CNOT operations, through observing the phase error information of the entanglement pair $\mathbb{2}$, we can classify entanglement pair $\mathbb{1}$ by two different phase error rates. One is
\begin{equation}
	\begin{split}
		&\tilde{e}_1^{\prime \mathrm{ph}}=\frac{(e^{\mathrm{ph}})^2}{(e^{\mathrm{ph}})^2+(1-e^{\mathrm{ph}})^2},
	\end{split}\label{eph1}
\end{equation}	
if there isn't any phase error in the entanglement pair $\mathbb{2}$ after the bilateral CNOT operations; the other is 
\begin{equation}
	\tilde{e}_2^{\prime \mathrm{ph}}=\frac{1}{2}\label{eph2},
\end{equation}
if there is a phase error in the entanglement pair $\mathbb{2}$. In the real protocol, all these entanglements and CNOT operations are not needed. We simply use Eq.~\eqref{eph1} and Eq.~\eqref{eph2} for phase error iteration after random pairing.

Based on the above analysis, we can obtain the asymptotic length $l$ of the key after error correction and privacy amplification in QKD with random pairing,
\begin{equation}
	\begin{split}\label{qkd}
		l=&N_t^{\prime}\big\{\Delta_{\mathrm{un}}^{\prime}-\Delta_{\mathrm{un}}^2\big[p_1H(\tilde{e}_1^{\prime\mathrm{ph}})+(1-p_1)H(\tilde{e}_2^{\prime\mathrm{ph}})\big]-\\
		&2\Delta_{\mathrm{un}}(1-\Delta_{\mathrm{un}})H(e^{\mathrm{ph}})\big\}-fN_t^{\prime}H(E^{\prime}),\\
	\end{split}
\end{equation}	
where  $H(x)=-x\log_2x-(1-x)\log_2(1-x)$ is the binary Shannon entropy function; $p_1=(e^{\mathrm{ph}})^2+(1-e^{\mathrm{ph}})^2$ and $f$ is the error correction coefficient and usually in the range of $1.1$ to $1.2$. 

\begin{remark}
The KGP of QDS does not need the steps of error correction and private amplification, because the definition of security in QDS is different from that of QKD~\cite{amiri2016secure,puthoor2016measurement}. 
\end{remark}

\begin{remark}
	The RP method in this work is to use the parity of two bits for one new outcome bit, there is no post-selection, no need of two-way classical communication. The method here is to reduce the phase-flip error while the bit-flip rises. The AOPP method in our earlier work~\cite{xu2020sending} takes parity check of two bits and then discard one bit and use the other bit for the surviving pair, discard both bits in the pair that has not passed the parity check. There is a post-selection and two-way classical communication is needed. The AOPP method there is to reduce the bit-flip error by iteration, while the phase-error there rises. 
\end{remark}

\section{Applications}\label{s3}

Our major idea of random pairing together with its iteration formulas Eqs.~(\ref{Ep}-\ref{eph2}) for error rates after random pairing applies for all QKD protocols for QDS task, here we take two important protocols to show the advantages of random pairing. One is the well known SNS~\cite{wang2018twin} protocol, which has been experimentally demonstrated by a number of experiments recently, including the first TF-QKD experiment with real setup~\cite{SNS_1realsetup}, the experiment over a secure distance of 509 km~\cite{SNS509}, which is the first TF-QKD exceeding a secure distance of 500 km with MDI security, the experiments over 550 km and 600 km~\cite{SNS600}, and very importantly, the field tests over 428 km~\cite{SNS428} and 511 km~\cite{SNS511}.
We shall show how to apply our random pairing method to the SNS protocol to achieve quantum digital signature, i.e., RP-SNS-QDS. Moreover, we shall also take our random pairing method to SCF-QKD proposed by Wang et al.~\cite{wang2019practical} for RP-SCF-QDS. Numerical simulation shows clearly the significant advantages of random pairing in QDS to improve the signature efficiency.

\begin{figure}
	\centering
	\includegraphics[scale=0.35]{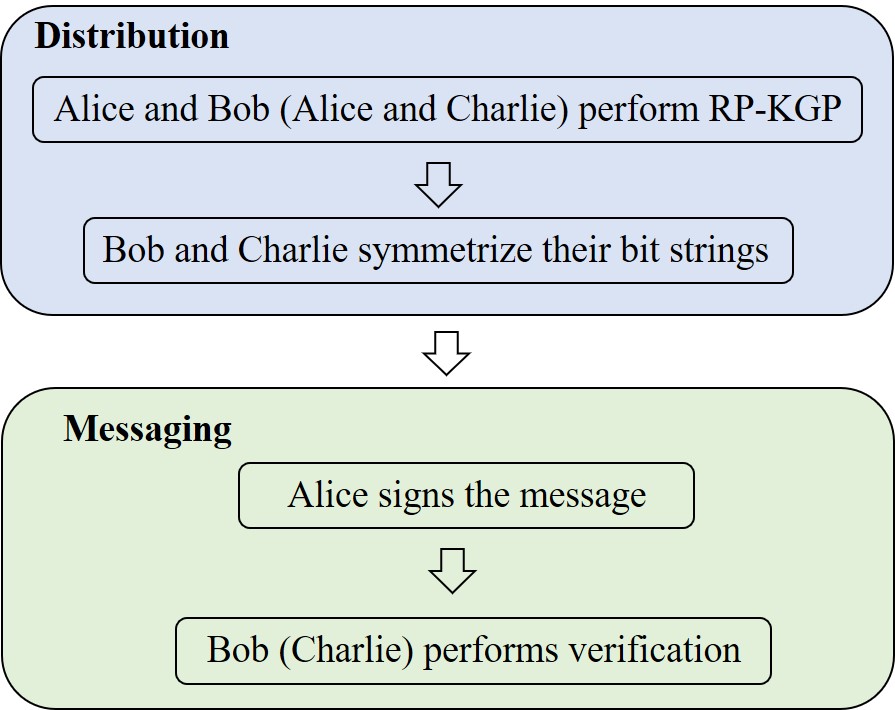}
	\caption{The flow chart of our RP-QDS protocol, which has two stages, the first is the distribution stage including RP-KGP and symmetrization; the second is the messaging stage, where the message is signed and verified. The details are introduced in the main text.}
	\label{QDS_sym}
\end{figure}

We show the flow chart of our RP-QDS protocol in Fig.~\ref{QDS_sym}. 

\subsection{RP-SNS-QDS}

\subsubsection{Distribution stage}

The RP-SNS-KGP of Alice-Bob is the same as that of Alice-Charlie and they can be done simultaneously. We take RP-SNS-KGP of Alice-Bob as an example.

(1) In each time window, Alice and Bob randomly choose the signal window with probability $p_z$ and the decoy window with probability $1-p_z$. 

In the signal window, Alice (Bob) decides to send Eve the phase-randomized coherent state of intensity $\mu$ with the probability of $q$ and writes down the classical bit value $1$ ($0$); Alice (Bob) decides to not send this phase-randomized coherent state, that is, send the vacuum state to Eve, with the probability of $1-q$ and writes down the classical bit value $0$ ($1$).
In the decoy window, Alice (Bob) randomly sends weak coherent state $\ket{\sqrt{\mu_k}e^{\imath \delta_k}}$, $k=1,2,\dots$, with probability $p_1, p_2, \dots,$ to Eve, where $\delta_k$ is random in $[0,2\pi)$.

(2) Eve uses a beam splitter to make measurements and announces the measurement outcome in each time window. 

\begin{Def}(Effective events, effective bits and effective windows in RP-SNS-QDS).
	Effective events are the events where one and only one detector  clicks. The bits and time windows corresponding to effective events are called effective bits and effective windows, respectively. 
\end{Def}

\begin{Def} (${Z}$ window, $\tilde{X}$ window in RP-SNS-QDS).
	The ${Z}$ window includes all events when both parties (Alice and Bob) choose the signal window. 
	If both parties choose the decoy window, it is an $\tilde{X}$ window. 
\end{Def}

\begin{Def} (Untagged bits in RP-SNS-QDS).
	The effective bits of single-photon states in $Z$ windows where one party (between Alice and Bob) decides to send and the other party decides not to send are untagged bits. 
\end{Def}

(3) Using the decoy state method, the proportion of untagged bits and the phase flip error rate can be estimated.
These effective bits used for parameter estimation are discarded.
After this, Alice and Bob obtain the associated bit strings $z_A$ and $z_B$, each one of length $N_t$, which are effective bits in $Z$ windows.
Instead of directly applying this data for QDS \cite{amiri2016secure,puthoor2016measurement,WangTFQDS2021}, here they make random pairing to generate outcome bit strings $z_A^{\prime}$ and $z_B^{\prime}$ for signature. They test the bit-flip error rate $E^{\prime}$ for their outcome bit strings. They also compute parameter values requested for QDS, such as phase flip error rate for the outcome bit strings. Details are shown later around~Eq.\eqref{mH}.    
The bit strings generated by RP-SNS-KGP between Alice and Bob forms outcome bit string with length of $N_t^{\prime}$. 

(4) Alice and Charlie perform the same steps above to generate a pair of associated bit strings for signature. 

(5) For each possible one-bit message $m=0$ or $1$, Alice and Bob (Alice and Charlie) select outcome bit strings for signature, which are $S_0^{B}$, $S_1^{B}$ for Alice and $K_0^{B}$, $K_1^{B}$ for Bob ($S_0^{C}$, $S_1^{C}$ for Alice and $K_0^{C}$, $K_1^{C}$ for Charlie), each one of length $\mathcal{L}$. The subscript $m$ of $S_m^p$ and $K_m^p$ represents the message $m$ that needs to be signed and the superscript $p\in \{B,C\}$ indicates that the bit string is generated by Alice in collaboration with the receiver Bob or Charlie.

(6) Bob and Charlie symmetrize their bit strings $K_m^B$ and $K_m^C$. Bob and Charlie randomly select $K_m^{B,\mathrm{s}}$ and $K_m^{C,\mathrm{s}}$ of length $\mathcal{L}/2$ and send these bit values and their corresponding positions to each other, respectively. The remaining bits $K_m^{B,\mathrm{k}}$ and  $K_m^{C,\mathrm{k}}$ of length $\mathcal{L}/2$ are kept locally. The superscript $\mathrm{s}$ of $K_m^{p,\mathrm{s}}$ represents the bits sent to each other and the superscript $\mathrm{k}$ of $K_m^{p,\mathrm{k}}$ represents the bits kept locally.
The symmetrized bit strings held by Bob and Charlie are recorded as $S_m^{\prime B }=(K_m^{B,\mathrm{k}},K_m^{C,\mathrm{s}})$ and $S_m^{\prime C}=(K_m^{B,\mathrm{s}},K_m^{C,\mathrm{k}})$ of length $\mathcal{L}$, respectively.

\subsubsection{ Messaging stage}

(1) Alice sends the message $m$ along with the signature $\mathrm{Sig}_m=\left(S_m^B,S_m^C\right)$ with length of $2\mathcal{L}$ to one desired receiver, for example, Bob.

(2) Bob determines whether to accept the message $m$ or not by comparing $\mathrm{Sig}_m$ with his bit string $S_m^{\prime B}$. If two parts, one is the comparison between $K_m^{B,\mathrm{k}}$ and the corresponding bits of $S_m^B$, and the other is the comparison between $K_m^{C,\mathrm{s}}$ and the corresponding bits of $S_m^C$, both have a mismatch rate less than $s_a$ ($<1/2$), then Bob accepts the message. Otherwise, Bob rejects the message.

(3) Bob forwards the $(m,\rm{Sig}_m)$ he received from Alice to Charlie.

(4) Charlie makes a similar comparison to determine whether to accept the message $m$ or not. If the mismatch rates of two parts, one is the comparison between $K_m^{B,\mathrm{s}}$ and the corresponding bits of $S_m^B$, and the other is the comparison between $K_m^{C,\mathrm{k}}$ and the corresponding bits of $S_m^C$, are less than $s_v$ ($0<s_a<s_v<1/2$), then Charlie accepts the message. Otherwise, Charlie rejects the message.

\subsection{RP-SCF-QDS}

The SCF-QDS protocol does not require switching physical bases to modulate the states differently. So it is not only immune to all attacks in the side-channel space of source-states but also retains the security of measurement-device-independence. 

The difference between RP-SCF-QDS and RP-SNS-QDS is the KGP in distribution stage.
Here we list the different steps. 
The RP-SCF-KGP of Alice-Bob is the same as that of Alice-Charlie and they can be done simultaneously. Here, we take the RP-SCF-KGP between Alice and Bob as an example.

(1) In each time window, Alice (Bob) prepares a coherent state $\ket{\alpha_A}$ ($\ket{\alpha_B}$) and announces it,   

\begin{equation}
	\begin{split}
		&\ket{\alpha_A}=e^{-\mu/2}\sum_{n=0}^{\infty}\frac{\mu^{n/2}}{\sqrt{n!}}e^{\imath\gamma_A}\ket{n},\\
		&\ket{\alpha_B}=e^{-\mu/2}\sum_{n=0}^{\infty}\frac{\mu^{n/2}}{\sqrt{n!}}e^{\imath\gamma_B}\ket{n},\\
	\end{split}
\end{equation}
where $\mu$ is the intensity of the coherent state; $\gamma_A$ ($\gamma_B$) is the global phase and $\imath$ is the imaginary unit.

Alice (Bob) decides to send Eve the coherent state $\ket{\alpha_A}$ ($\ket{\alpha_B}$) with the probability of $q$ and writes down the classical bit value $1$ ($0$); Alice (Bob) decides to not send the coherent state $\ket{\alpha_A}$ ($\ket{\alpha_B}$), that is, sends the vacuum state to Eve, with the probability of $(1-q)$ and writes down the classical bit value $0$ ($1$).

(2) Eve uses a beam splitter to make measurements and announces the measurement outcome in each time window. The definitions of effective event, effective bits and effective windows are the same as those in RP-SNS-QDS.

\begin{Def} ($\tilde{Z}$ window, $\mathcal{B}$ window and $\mathcal{O}$ window in RP-SCF-QDS).
	A time window when one party (between Alice and Bob) decides to send the coherent state and the other party decides to not send  the coherent state is $\tilde{Z}$ window. A time window when both parties decide to send the coherent state or when both parties decide to not send the coherent state is 
	$\mathcal{B}$ window or $\mathcal{O}$ window.
\end{Def}

\begin{Def} (Untagged bits in RP-SCF-QDS).
	The effective bits in $\tilde{Z}$ window are untagged bits. 
\end{Def}

(3) In all time windows, Alice and Bob choose two random subsets $v$ and $u$ through classical communication. The subset $v$ is used to test the bit-flip error rate. They can use the data in the subset $v$ to estimate the bounds of some quantities in the subset $u$ to obtain the upper bound of the phase-flip error rate. These efficient bits used for parameter estimation are discarded. 
After Alice and Bob obtain the associated bit strings with the length of $N_t$, they make random pairing and use the parity values of the bit pairs as the new bit string for signature. They test the bit-flip error rate $E^{\prime}$ for their outcome bit strings and calculate phase flip error rate for the outcome bit strings. Details are shown later around~Eq.\eqref{mH}.
The bit strings generated by RP-SCF-KGP have the length of $N_t^{\prime}$. 

The step (4)-(6) of distribution stage and the step (1)-(4) of the messaging stage are the same as those in RP-SNS-QDS.

We use the quantum communication of SCF-QKD ~\cite{wang2019practical} to generate the associated bit strings for signature and the subsequent steps of RP-SCF-QDS can be regarded as the post data processing. By adopting the security analysis method of SCF-QKD \cite{wang2019practical}, we can also demonstrate that our RP-SCF-QDS is immune to all attacks in the side-channel space of emitted photons and measurement-device independent. 

\subsection{Others}

Straightly, the RP method can also be applied for  decoy-state MDIQKD \cite{WangThreeMDI2013,zhouFourMDI2016} and decoy-state BB84 \cite{Hwang2003,Wangdecoy2005,Lodecoy2005,PhysRevLett.121.190502} protocol: after quantum communication stage of this protocols, we can find the bit-flip error rate, verify the single-photon phase-flip error rate and the number of untagged bits by decoy-state analysis. We then calculate the single-photon phase-flip rate and number of untagged after RP by iteration formulas Eqs.~\eqref{Ep}-\eqref{eph2} and then QDS can be done with these. The bit-flip error rate after RP can be actually tested directly in a real protocol, though it can also be calculated.

It should be noted that the step of random pairing can be moved to messaging stage of our protocol. That is, there is no change in distribution stage, Alice and Bob (Alice and Charlie) just perform KGP in the standard way~\cite{amiri2016secure}: Alice sends out the bits from quantum communication of QKD (without random pairing) to generate associated bit strings and Bob and Charlie symmetrize their bit strings.  In messaging stage, Bob (Charlie) performs random pairing for the bit strings sent by Alice and the bit strings held locally. Bob (Charlie) determines whether to accept the message based on the comparison results of the outcome bit strings. From this point of view, the method of random pairing can provide a tightened result for the security level of QDS. For conciseness, here we shall only consider the method introduced earlier before Eq.~\eqref{Nt}, where Alice did the random pairing in distribution stage.

To test security, we need bound values of a number of parameters. In what follows, we will show how to work out all these values reliably and make a quantitative security analysis.


\section{Security Analysis}\label{s4}

The security level of quantum digital signature protocol is $ \varepsilon $ ~\cite{amiri2016secure}, if
\begin{equation}
	\varepsilon=\max\{P_{\mathrm{ro}},P_{\mathrm{fo}},P_{\mathrm{re}}\},
\end{equation} 
where $P_{\mathrm{ro}}$ represents the probability of an honest run aborting; $P_ {\mathrm{fo}} $ and $P_{\mathrm{re}}$ represent the probability of forging and repudiation, respectively.

\textit{Robustness.} Bob rejects a message that is signed by Alice when at least one  mismatch rate is greater than $s_a$ (one is the comparison between $K_m^{B,\mathrm{k}}$ and the corresponding bits of $S_m^B$, and the other is the comparison between $K_m^{C,\mathrm{s}}$ and the corresponding bits of $S_m^C$). Therefore, $P_{\mathrm{ro}}$ is associated with the error rate of parameter estimation of two parts, one ($\varepsilon_{1}$) is related to the outcome bit string generated by RP-KGP between Alice and Bob, the other ($\varepsilon_{2}$) is related to the outcome bit string generated by RP-KGP between Alice and Charlie,
\begin{equation}
	P_{\mathrm{ro}}\leq \varepsilon_{1}+\varepsilon_{2}.
\end{equation}

In the asymptotic case, $\varepsilon_1=\varepsilon_2=0$, thus $P_{\mathrm{ro}}=0$.

\textit{Unforgeability.} Internal parties (Bob and Charlie) are more likely to succeed in forging signatures than external parties.
If one receiver (Bob) generates a signature that is not received from Alice, but is accepted by another receiver (Charlie), then it means a successful forgery. Based on \cite{amiri2016secure}, in asymptotic case, the probability of forgery is

\begin{equation}
	\begin{split}
		& P_{\mathrm{fo}}\leq p_{\mathrm{F}}+g,\\
	\end{split}
\end{equation}
in which
\begin{equation}
	\begin{split}
		&p_F=\frac{1}{g}\{2^{-\frac{\mathcal{L}}{2}[\mathcal{H}-H(\frac{2r}{\mathcal{L}})]}+\varepsilon_e\},\\
	\end{split}
\end{equation}
and
\begin{equation}\label{mH}
	\begin{split}
		\mathcal{H}=&\Delta_{\mathrm{un}}^{\prime}-\Delta_{\mathrm{un}}^2\big[p_1H(\tilde{e}_1^{\prime\mathrm{ph}})+(1-p_1)H(\tilde{e}_2^{\prime\mathrm{ph}})\big]-\\
		&2\Delta_{\mathrm{un}}(1-\Delta_{\mathrm{un}})H(e^{\mathrm{ph}}),\\
	\end{split}
\end{equation}
where $\tilde{e}_1^{\prime\mathrm{ph}}$ and $\tilde{e}_2^{\prime\mathrm{ph}}$ are shown in~Eq.\eqref{eph1} and ~Eq.\eqref{eph2}, $g$ is the upper bound of probability that Bob makes fewer than $r$ errors in $K_{m}^{C,k}$ with length of $\mathcal{L}/2$ except with probability at most $p_F$; $\mathcal{H}$ is obtained from the analysis in section~\ref{s2}; $\varepsilon_e$ is related to the smooth min-entropy, which is shown in Eq.~\eqref{H}.

\textit{Transferability (or Nonrepudiation).}  The transferability of message means that if it is accepted by one honest receiver, then it will be accepted by another honest receiver. That is, if both honest receivers accept the message, then the signer cannot repudiate the signature. The steps of symmetrization in distribution stage can ensure the security against repudiation. According to \cite{amiri2016secure}, the probability of repudiation is

\begin{equation}
	P_{\mathrm{re}}\leq 2e^{-\frac{1}{4}(s_v-s_a)^2\mathcal{L}},
\end{equation}
where $s_a$ and $s_v$ are thresholds for the two receivers to decide whether to accept the message or not. 

\textit{Thresholds $s_a$ and $s_v$.}
The thresholds $s_a$ and $s_v$ are determined by the bit-flip error rate $E_{\mathcal{L}/2}$ of $K_m^{p,\mathrm{k}}$ with length of $\mathcal{L}/2$ and the minimum error rate $P_e$ when the eavesdropper (Bob or Charlie) guesses $K_m^{p,\mathrm{k}}$ ($p\in\{B,C\}$, $m\in\{0,1\}$),
\begin{equation}
	\begin{split}
		& s_a = E_{\mathcal{L}/2}+\frac{1}{3}(P_e-E_{\mathcal{L}/2}), \\
		& s_v = E_{\mathcal{L}/2}+\frac{2}{3}(P_e-E_{\mathcal{L}/2}). \\
	\end{split}
\end{equation}	

Either Bob or Charlie could be a malicious participant, for who the only unknown bit string is the one ($K_m^{p,k}$ with length of $\mathcal{L}/2$) that has not been sent in symmetrization step of distribution stage. For example, in order to forge a signature for Charlie to accept the message, Bob can be the eavesdropper.

In Supplemental Information, we show how to get the bit-flip error rate before random pairing. Combined with the iteration relations, the bit-flip error rate $E_{\mathcal{L}/2}$ after random pairing can be obtained. 
In the following, we analyze how to obtain $P_e$.

We can use the eavesdropper's smooth min-entropy to describe the average probability of guessing $K_m^{p,\mathrm{k}}$ within a certain threshold 
\begin{equation}\label{H}
	\begin{split}
		H_{\min}^{\varepsilon_e}(K_m^{p,k}|I_e)
		\ge
		&\frac{\mathcal{L}}{2}\mathcal{H},\\
	\end{split}
\end{equation}	
where $\mathcal{H}$ is shown in Eq.~\eqref{mH} and $I_e$ represents all information of the eavesdropper. Based on~\cite{amiri2016secure}, the minimum error rate $P_e$ when the eavesdropper guesses $K_m^{p,\mathrm{k}}$ can be obtained by

\begin{equation}
	\begin{split}
		&H(P_e)=\mathcal{H},\\
	\end{split}
\end{equation}
where $\mathcal{H}$ is shown in Eq.~\eqref{mH}.

\textit{Summary of parameters.} Alice sends $N$ pulses in total during RP-KGP with Bob. After the step of random pairing, Alice can generate the outcome bit strings with length of $N_t^{\prime}$. In the symmetric case, Alice does the same procedure with Charlie and generates the  $N_t^{\prime}-$bit string for signature.
For each possible one-bit message $m=0$ or $1$, Alice's signature is $\mathrm{Sig}_0=(S_0^{B},S_0^{C})$ or $\mathrm{Sig}_1=(S_1^{B},S_1^{C})$, each one of length $2\mathcal{L}$. That is, to sign $n_s$ one-bit messages, where $n_s=\frac{N_t^{\prime}}{2\mathcal{L}}$, Alice needs to send $N$ pluses during RP-KGP with Bob.  
So, the signature rate~\cite{WangTFQDS2021} $R$ of our RP-QDS protocol can be defined as  
\begin{equation}
	R = \frac{n_s}{N}.
\end{equation}
Given system parameters $(N,\alpha, \eta_d,P_d,g,\varepsilon)$, we should optimise over all other parameters to obtain the optimal signature rate $R$. Notation $\alpha$ is the loss coefficient of fiber, $\eta_d$ is the detection efficiency, $P_d$ is the dark count rate of detectors, the meaning of $g$ is explained following Eq.~\eqref{mH} and $\varepsilon$ is the security level. 

We show how to obtain the bit-flip error rate $E$ and phase-flip error rate $e^{\mathrm{ph}}$ of the bit string before random pairing in Supplemental Information. Combined with the iteration relations shown in Eqs.~\eqref{Ep}-\eqref{eph2}, the bit-flip error rate and phase-flip error rate of the outcome bit string can be obtained, which are used to determine the security level $\varepsilon$ and thresholds $s_a$, $s_v$ of our protocol.

\textit{Numerical simulations.}

\begin{table}[]
	\vspace{20pt}
	\centering
	\caption{System parameters used in numerical simulation.}
	\begin{tabular}{p{2cm}p{1.4cm}p{1.4cm}p{1.4cm}p{1.4cm}}
		\hline
		\hline
		$ \alpha $  & $ \eta_d $ & $ P_d $ & $g$ & $\varepsilon$ \\
		\hline
		$ 0.2\ \mathrm{dB/km} $ & $ 80\%$ &$ 10^{-11} $ & $10^{-12}$ & $10^{-5}$  \\
		\hline
		\hline
	\end{tabular}
	\label{t_p}
\end{table}

\begin{figure}
	\centering
	\includegraphics[scale=0.35]{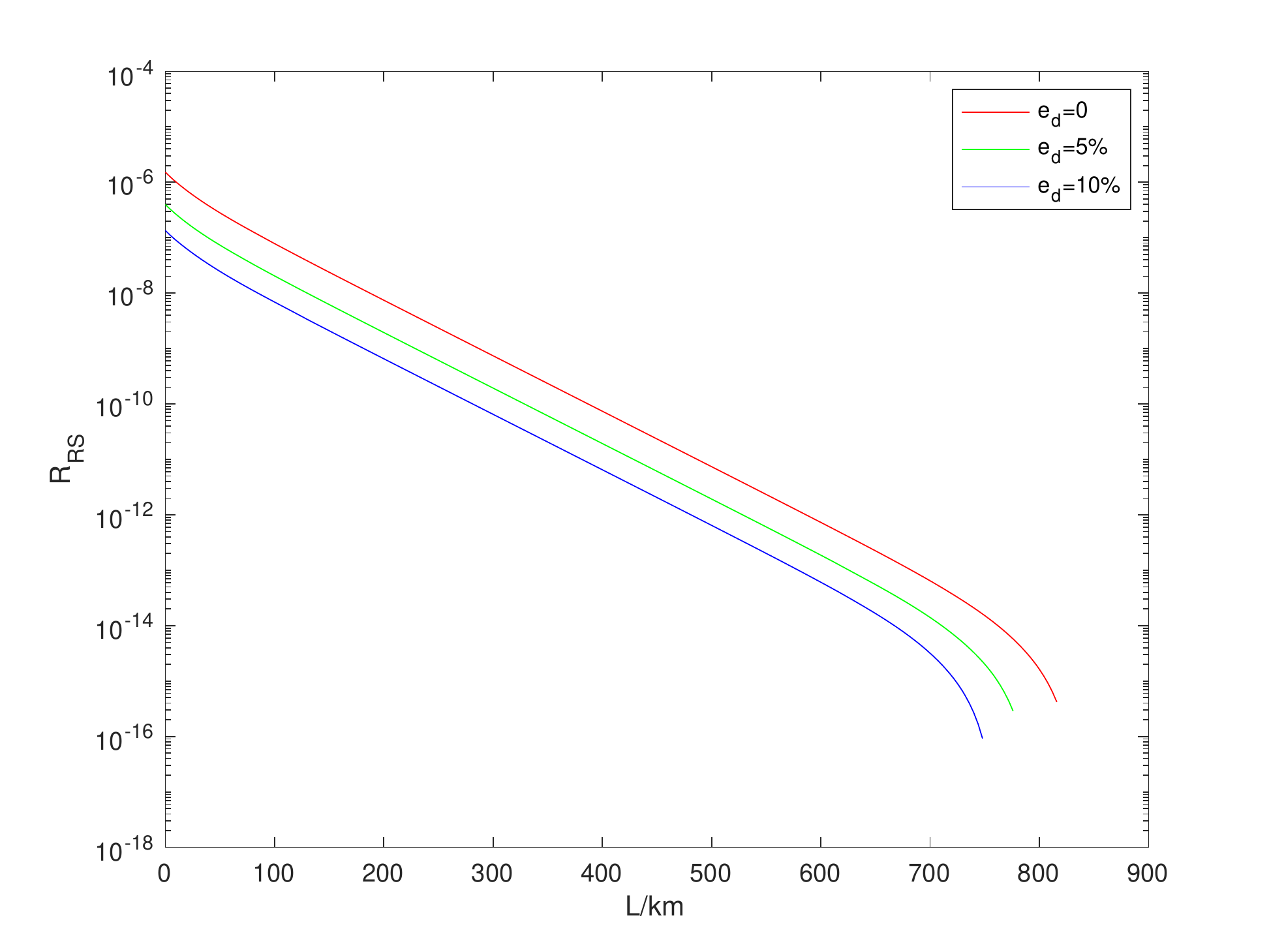}
	\caption{The red, green and blue lines represent the signature rates $R_{\mathrm{RS}}$ of RP-SNS-QDS under misalignment error rate $e_d=0$, $5\%$ and $10\%$, respectively.}
	\label{SNS_R}
\end{figure}

\begin{figure}
	\centering
	\includegraphics[scale=0.35]{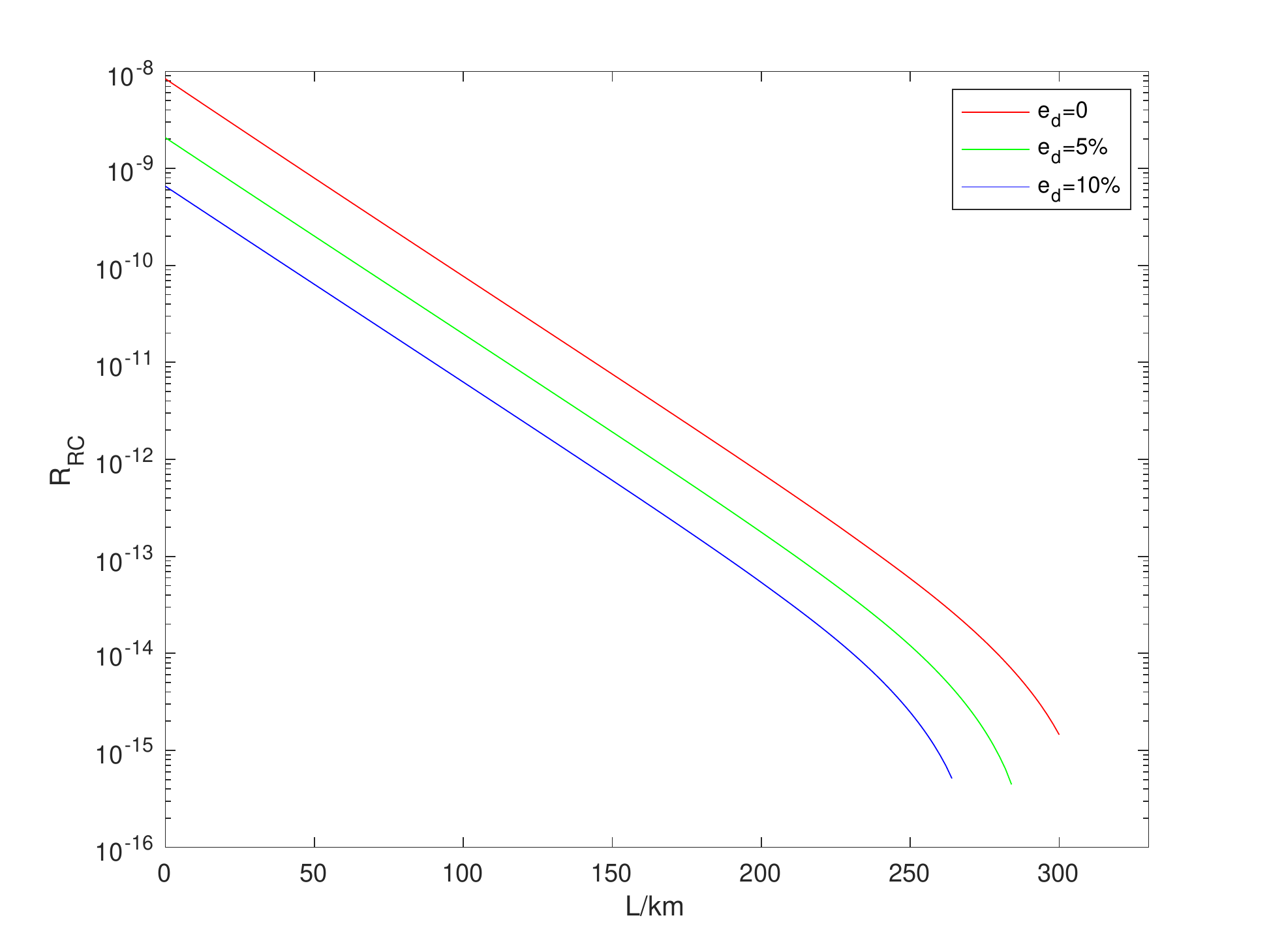}
	\caption{The red, green and blue lines represent the signature rates $R_{\mathrm{RC}}$ of RP-SCF-QDS under misalignment error rate $e_d=0$, $5\%$ and $10\%$, respectively.}
	\label{SCF_R}
\end{figure}

In Fig.~\ref{SNS_R} and Fig.~\ref{SCF_R}, we show the log scale of the signature rate $R_{\mathrm{RS}}$ (of RP-SNS-QDS) and $R_{\mathrm{RC}}$ (of RP-SCF-QDS) as a function of the distance $L$ (km)  between Alice and Bob, which is the same as the distance between Alice and Charlie in the symmetric case, under different misalignment error rates $e_d$. System parameters used in numerical simulation are shown in table~\ref{t_p}. For both protocols, with the increase of misalignment error rate $e_d$, the signature rate $R_{\mathrm{RS}}$ and $R_{\mathrm{RC}}$ decrease. It should be emphasized that the results in Fig.~\ref{SCF_R} have a higher level security, that is, our RP-SCF-QDS is not only measurement-device independent, but also immune to all attacks in the side-channel of emitted photons.

\begin{figure}
	\centering
	\includegraphics[scale=0.33]{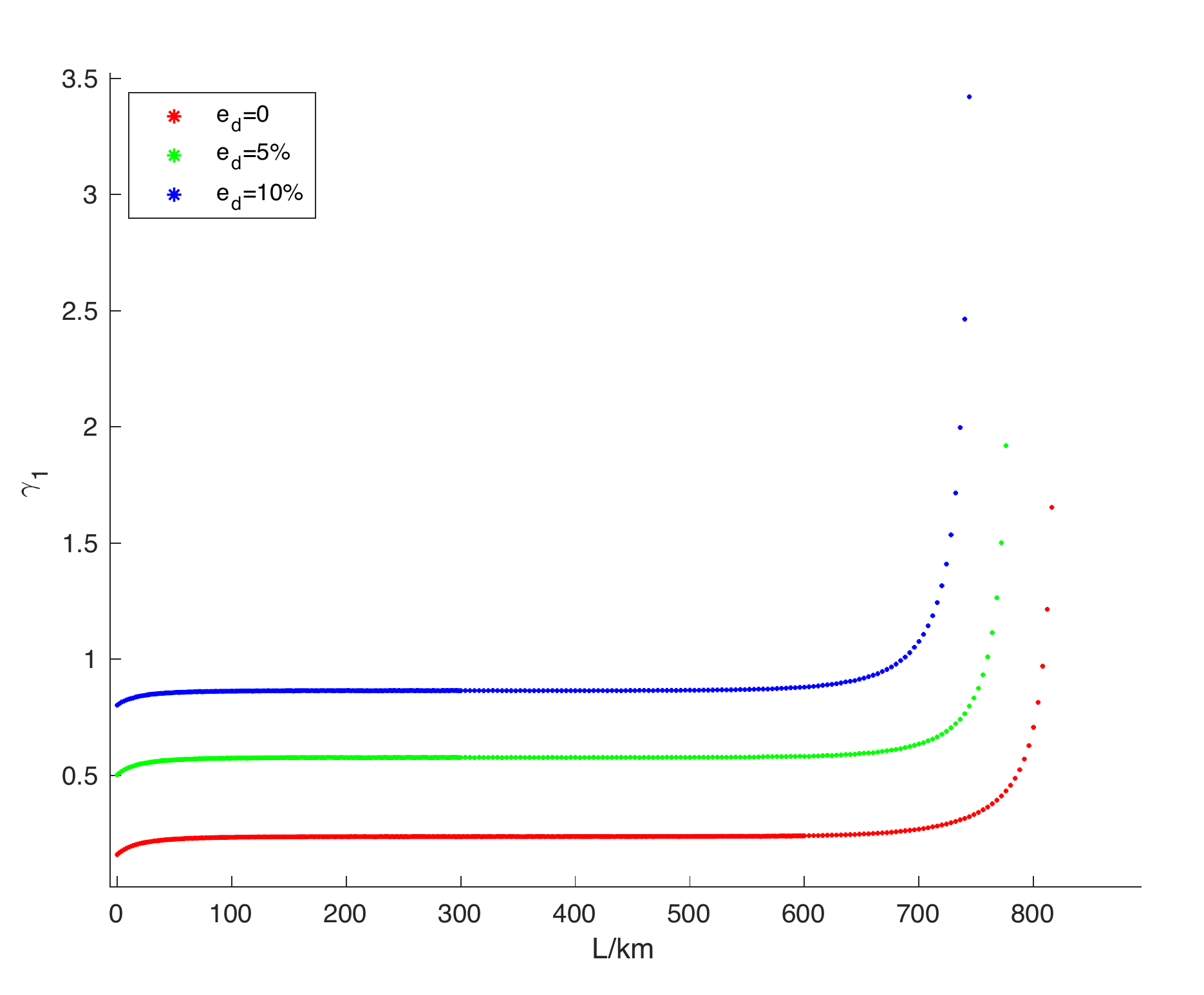}
	\caption{The signature rate can be increased by $\gamma_1$ with random pairing using SNS-QDS protocol under misalignment error $e_d=0$ (red), $5\%$ (green) and $10\%$ (blue).}
	\label{p_i1}
\end{figure}

\begin{figure}
	\centering
	\includegraphics[scale=0.33]{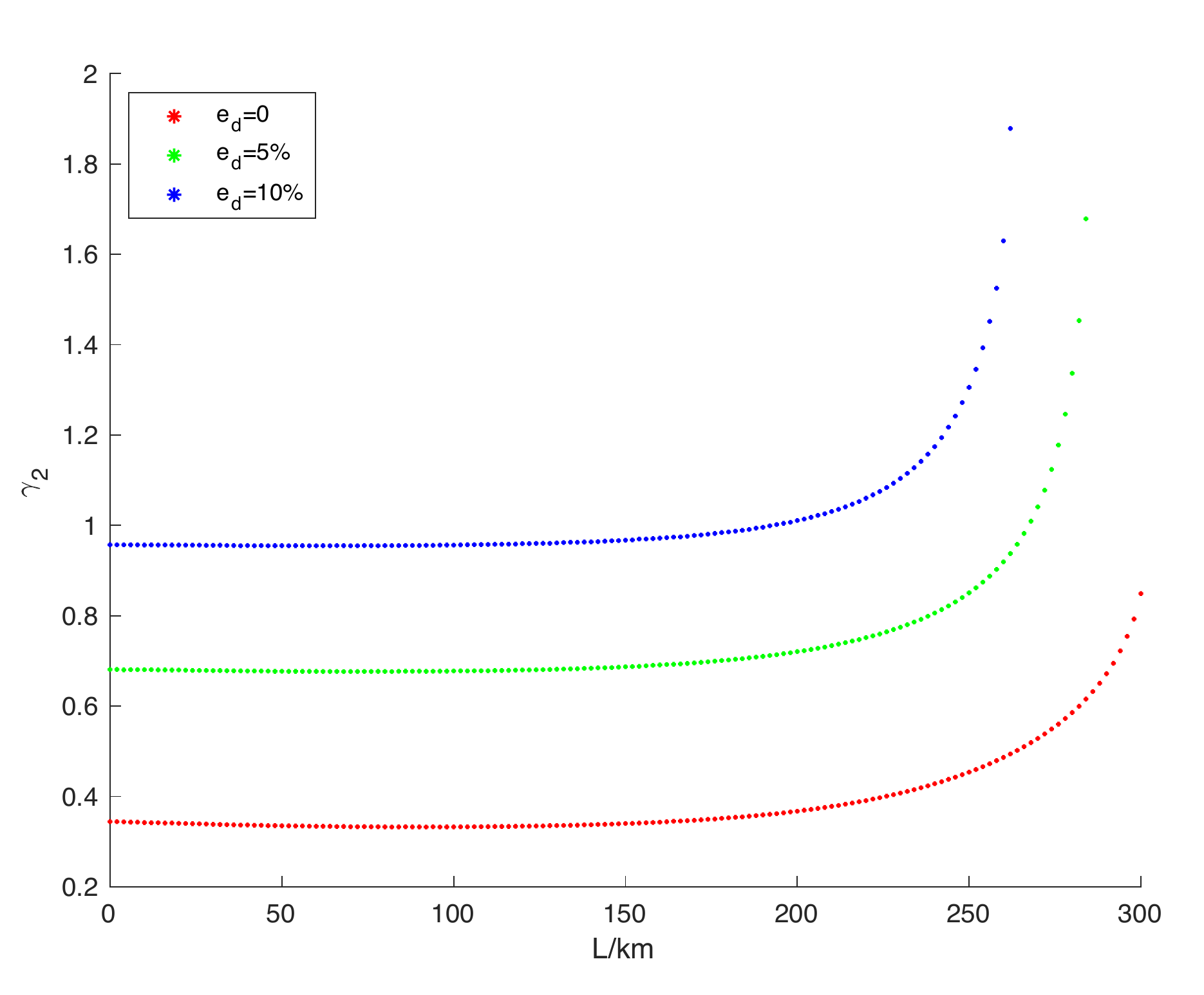}
	\caption{The signature rate can be increased by $\gamma_2$ with random pairing using SCF-QDS protocol under misalignment error $e_d=0$ (red), $5\%$ (green) and $10\%$ (blue).}
	\label{p_i2}
\end{figure}

In Fig.~\ref{p_i1} and  Fig.~\ref{p_i2}, we show that the signature rate can be increased by $\gamma_1$ and $\gamma_2$ with random pairing using SNS-QDS and SCF-QDS, respectively, where $\gamma_1=\frac{R_{\mathrm{RS}}}{R_{\mathrm{S}}}-1$, $\gamma_2=\frac{R_{\mathrm{RC}}}{R_{\mathrm{C}}}-1$; $R_{\mathrm{S}}$ and $R_{\mathrm{C}}$ are signature rate of SNS-QDS and SCF-QDS.
It can be found that for both protocols, applying random pairing can greatly improve the signature rate, that is, $\gamma_1>0$ and $\gamma_2>0$, and this advantage becomes more and more obvious with the increase of misalignment error $e_d$. For instance, when $L=200$ km, $e_d=10\%$, the signature rate of SNS-QDS and SCF-QDS can be increased by $86\%$ and $101\%$ with random pairing, respectively.

\section{Random pairing with finite pluses}\label{s5}

In any real application of quantum cryptography, the data size is always finite. Effects of finite data size has to be robustly considered for security. Here we shall optimize the signature rate with finite data size.  

We focus on signing a one-bit message $m=0$ or $1$. Alice does RP-KGP with Bob, during which she sends $N_f$ pulses in total and generates the outcome bit string with length of $N_{t,f}^{\prime}$. In the symmetric case, Alice does the same procedure with Charlie and generates the   $N_{t,f}^{\prime}-$bit string. That is, to sign a one-bit message ($n_s=1$), Alice needs to send $N_f$ pluses in total during RP-SNS-KGP with Bob.  
So, the signature rate \cite{WangTFQDS2021} $R_f$ is
\begin{equation}
	R_f = \frac{1}{N_f}.
\end{equation}
Our practical RP-SNS-QDS protocol can be extended to sign multi-bit messages using the method in \cite{wang2015longmessages}. The main idea of \cite{wang2015longmessages} is to encode each bit 0 (1) of the multi-bit message into 000 (010) and add codeword 111 to the start and the end of the message. This approach improves the security of signing multi-bit messages, which can resist more attacks on multi-bit messages, but has no effect on the advantages of random pairing. 

All parameters values listed above for the optimization of signature rate given system parameters have to be estimated robustly with finite data size. Here we show the estimation for bit-flip error rate and phase-flip error rate of outcome bit string. The other parameter estimations are presented in Supplemental Information. Firstly, to obtain the bit-flip error rate $E^{\prime}_f$ of $K_m^{p,k}$ with length of $\mathcal{L}/2$, $T$ bits are randomly selected and we use the bit-flip error rate $E_T$ of these $T$ bits to estimate $E^{\prime}_f$ \cite{Serfling1974},
\begin{equation}\label{E}
	\begin{split}
		&E^{\prime}_f \le E_T+\mu(\frac{\mathcal{L}}{2},T,\varepsilon_{\mathrm{PE}}),\\
	\end{split}
\end{equation}
where
\begin{equation}
	\begin{split}
		&   \mu(\frac{\mathcal{L}}{2},T,\varepsilon_{\mathrm{PE}})=\sqrt{\frac{(\frac{\mathcal{L}}{2}-T+1)\ln(\frac{1}{\varepsilon_{\mathrm{PE}}})}{T\mathcal{L}}},\\
	\end{split}
\end{equation}
in which except for a small probability $\varepsilon_{\mathrm{PE}}$, the above estimation is successful. As analyzed above, the bit-flip error rate $E_T$ for thees $T$ outcome bits can be tested directly. When only one of the paired bits in original bit string has an error, the new bit made up of their parity value will have an error. So the expected value for $E_T$ is
\begin{equation}\label{fe}
	E_T = 2E(1-E),
\end{equation}
where $E$ is the bit-flip error rate of original bit strings.	
It should be noted that $E_T$ can be directly observed in experiments and we use $E_T$~\eqref{fe} in numerical simulations. 

Then we analyze how to obtain the phase-flip error rate of $K_m^{p,k}$. There are two cases: one is that the outcome bit is generated by two untagged bits, and the other is that the outcome bit is generated by one tagged bit and one untagged bit. 
For the first case (two untagged bits are paired), we classify the outcome bits according to the number of phase errors carried by the bit pairs.
Notation $n_x$, $x\in\{0,1,2\}$, represents the number of bit pairs consisting of two untagged bits with $x$ phase error(s) in $K_m^{p,k}$ with length of $\frac{1}{2}\mathcal{L}$,

\begin{equation}\label{n0}
	n_0 = \frac{1}{2}\mathcal{L}\Delta_{u,f}^2(1-e^{\mathrm{ph}})^2, 
\end{equation}	

\begin{equation}\label{n1}
	n_1 = \mathcal{L}\Delta_{u,f}^2e^{\mathrm{ph}}(1-e^{\mathrm{ph}}), 
\end{equation}	

\begin{equation}\label{n2}
	n_2 =\frac{1}{2}\mathcal{L}\Delta_{u,f}^2(e^{\mathrm{ph}})^2, 
\end{equation}
where $\Delta_{u,f}$ and $e^{\mathrm{ph}}$ is the proportion of untagged bits and the phase-flip error rate of the original bit string, respectively. 
When the total number of phase error is even ($0$ and $2$ phase error(s)), the corresponding phase-flip error rate is $\frac{n_2}{n_0+n_2}$. When the total number of phase error is odd ($1$ phase error), the corresponding phase-flip error rate is $\frac{1}{2}$, which has no contribution to the signature.

For the second case (one tagged bits and one untagged bit are paired), the phase-flip error rate is $e^{\mathrm{ph}}$ and the number of bit pairs in $K_m^{p,k}$ with length of $\frac{1}{2}\mathcal{L}$ is
\begin{align}\label{nut}
	n_{\mathrm{ut}} = \mathcal{L}\Delta_{u,f}(1-\Delta_{u,f}).
\end{align}	 

The smooth min-entropy $H_{\min}^{\varepsilon_e}(K|I_e)$ can be used to characterize the average probability when the forger guesses $K_m^{p,k}$ within a threshold, 

\begin{equation}\label{}
	\begin{split}
		H_{\min}^{\varepsilon_e}(K_m^{C,k}|I_e)
		\ge
		&\frac{1}{2}\mathcal{L}\mathcal{H}_f,\\
	\end{split}
\end{equation}	
where $I_e$ represent all information of the forger, and

\begin{equation}\label{hf}
	\begin{split}
		\mathcal{H}_f=&\frac{2}{\mathcal{L}}\big[(n_{0}^{\mathrm{L}}+n_{2}^{\mathrm{U}})[1-H(\frac{n_2^{\mathrm{U}}}{n_0^{\mathrm{L}}+n_2^{\mathrm{U}}})]\\
		&+n_{\mathrm{ut}}^{\mathrm{L}}[1-H(e^{\mathrm{ph}})]\big],\\
	\end{split}
\end{equation}
where $V^{\mathrm{L}}$ ($V^{\mathrm{R}}$) represents the lower bound (upper bound) of $V$ and the calculation of $\mathcal{H}_f$ is explained in detail in Supplemental Information .

Considering the finite data size effects, the probabilities $P_{\mathrm{ro}},P_{\mathrm{fo}},P_{\mathrm{re}}$ related to the security level $\varepsilon$ ($=\max\{P_{\mathrm{ro}},P_{\mathrm{fo}},P_{\mathrm{re}}\}$) become 
\begin{equation}
	P_{\mathrm{ro},f}\leq 2\varepsilon_{\mathrm{PE}},
\end{equation}
where $\varepsilon_{\mathrm{PE}}$ is the failure probability of parameter estimation, which is shown in Eq.~\eqref{E},
\begin{equation}
	\begin{split}
		& P_{\mathrm{fo},f}\leq \frac{1}{g}\{2^{-\frac{\mathcal{L}}{2}[\mathcal{H}_f-H(\frac{2r}{\mathcal{L}})]}+\varepsilon_e\}+g+\varepsilon_{\mathrm{PE}}+8\xi,\\
	\end{split}
\end{equation}
where $g$ is the upper bound of probability that Bob makes fewer than $r$ errors in $K_{m}^{C,k}$ except with a small probability of $\frac{1}{g}\{2^{-\frac{\mathcal{L}}{2}[\mathcal{H}-H(\frac{2r}{\mathcal{L}})]}+\varepsilon_e\}$; $8\xi$ is generated in parameter estimating (the detail is shown in Supplemental Information ); $\varepsilon_e$ is related to the smooth min-entropy, which is shown in Eq.~\eqref{mH},

\begin{equation}
	P_{\mathrm{re},f}\leq 2e^{-\frac{1}{4}(s_v-s_a)^2\mathcal{L}}.
\end{equation}

\begin{figure}
	\centering
	\includegraphics[scale=0.48]{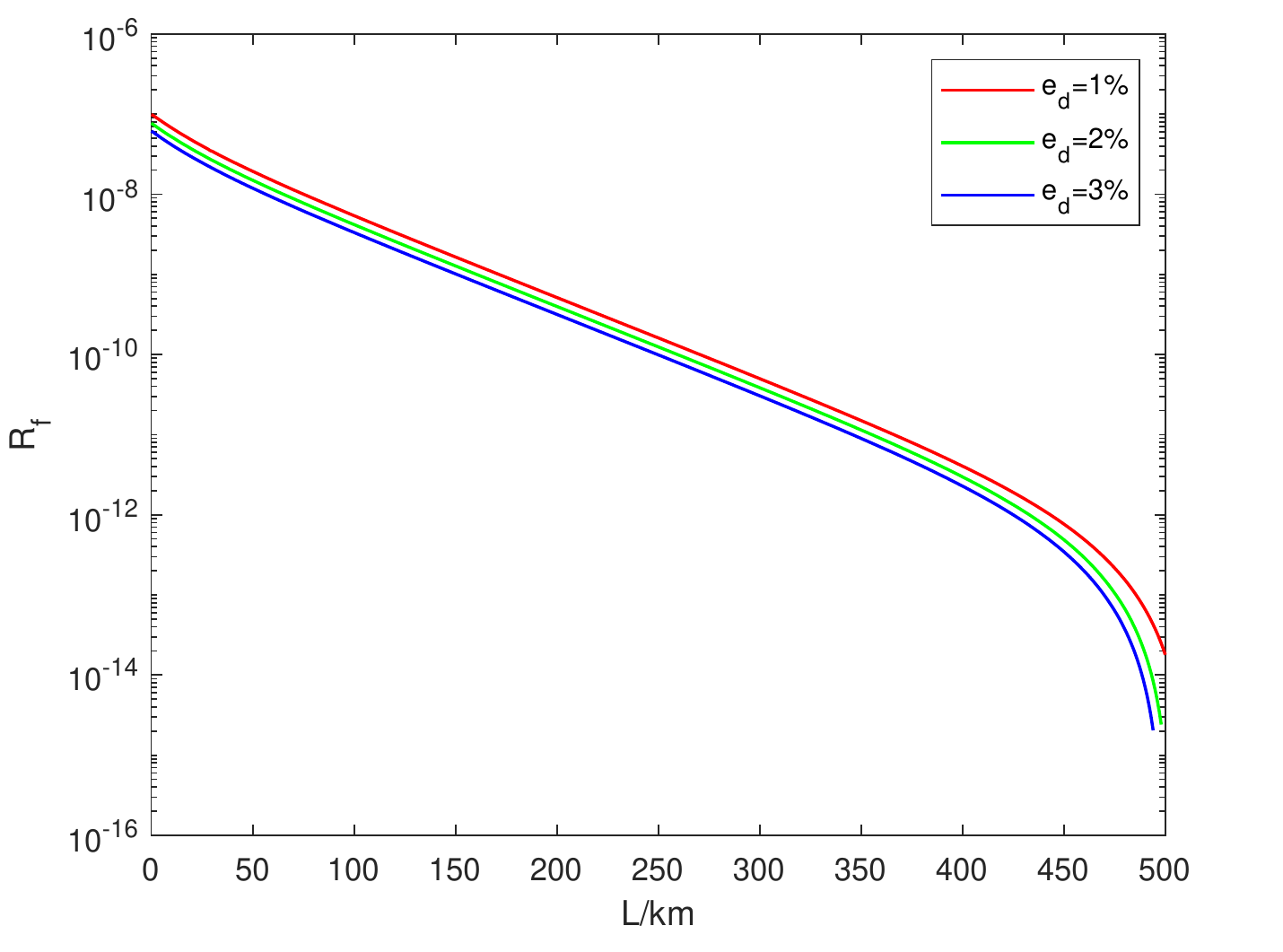}
	\caption{The red, green and blue lines represent the signature rates $R_f$ of RP-SNS-QDS with finite-size effects under misalignment error rate $e_d=1\%$, $2\%$ and $3\%$, respectively, where $\varepsilon = 10^{-5}$, $\alpha=0.2$ dB/km, $\eta_d=80\%$ and $p_d=10^{-8}$.}
	\label{pN}
\end{figure}

Given system parameters $(\alpha, \eta_d,P_d,\varepsilon)$, we need to optimize the signature rate $R_f$ in RP-SNS-QDS with finite-size effects, that is, maximize $R_f$ by choosing appropriate parameters $(\mu,q,\mu_1,\mu_2,p_z,p_0,\Delta,\gamma_T,\varepsilon_{PE},g,\xi)$, which are detailedly analyzed in Supplemental Information. Without loss of generality, we consider the symmetric case, where the distance between every participant (Alice, Bob and Charlie) and Eve is the same. In Fig.~\ref{pN}, we show the signature rate $R_f$ as a function of the distance $L$ (km) between Alice and Bob when misalignment error rate $e_d\in\{1\%,2\%,3\%\}$ in our practical RP-SNS-QDS. Obviously, as $e_d$ increases, the signature rate $R_f$ decreases.

\begin{figure}
	\centering
	\includegraphics[scale=0.45]{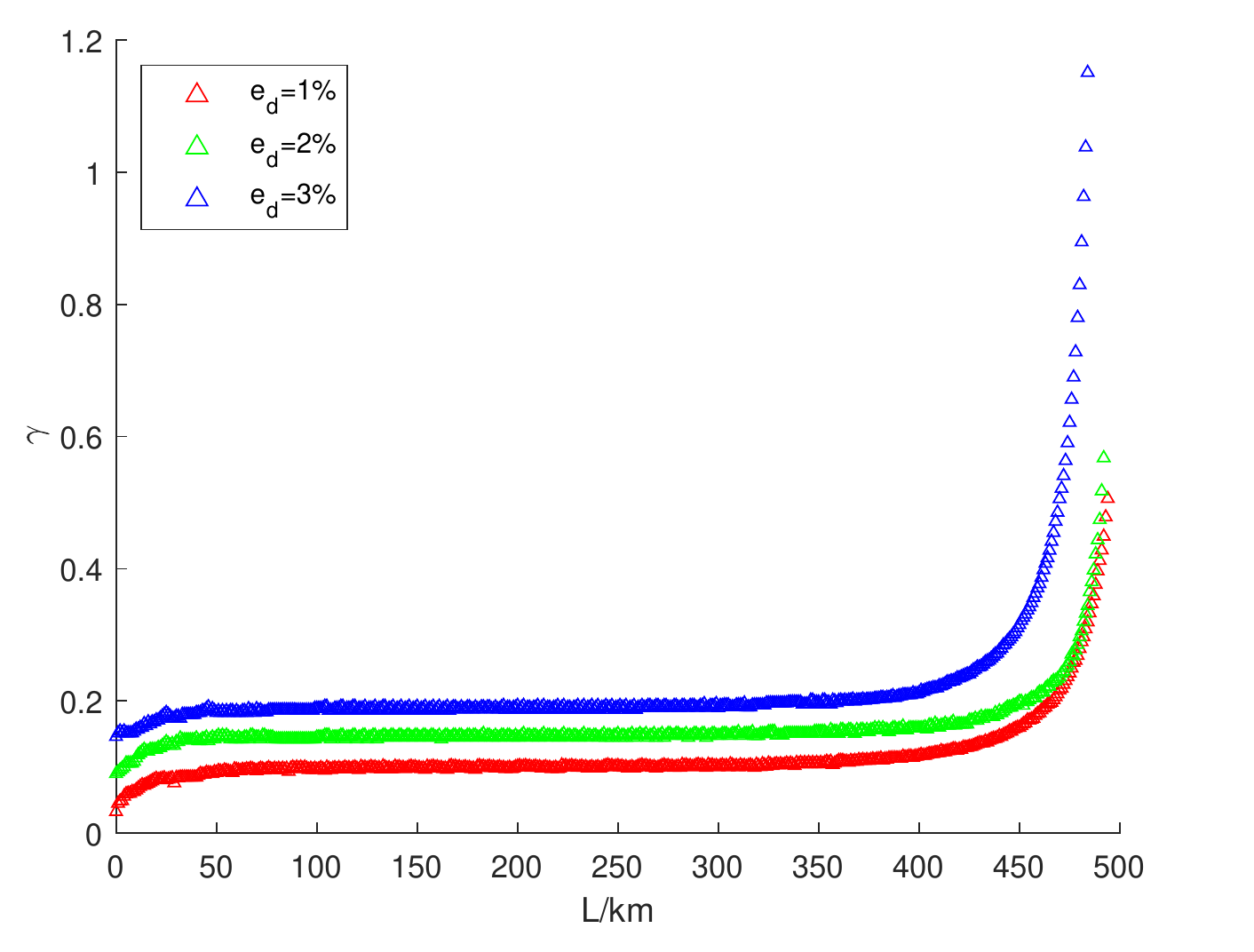}
	\caption{The signature rate can be increased by $\gamma$ with random pairing in our practical SNS-QDS compared with the one without random pairing \cite{WangTFQDS2021} under misalignment error $e_d=1\%$ (red), $2\%$ (green) and $3\%$ (blue).}
	\label{pgamma}
\end{figure}

In Fig.~\ref{pgamma}, we show that the signature rate can be increased by $\gamma$ with random pairing in our protocol, where $\gamma=\frac{R_f}{R_o}-1$, $R_o$ and $R_f$ are signature rate of SNS-QDS and our practical RP-SNS-QDS. For example, when $e_d=3\%$, $L=483$ km, using random pairing, the signature rate can be increased by about $104\%$. It can be found that as the distance $L$ and $e_d$ increases, the advantage of random pairing becomes more and more obvious.

\section{Discussion}
\begin{figure}
	\centering
	\includegraphics[scale=0.48]{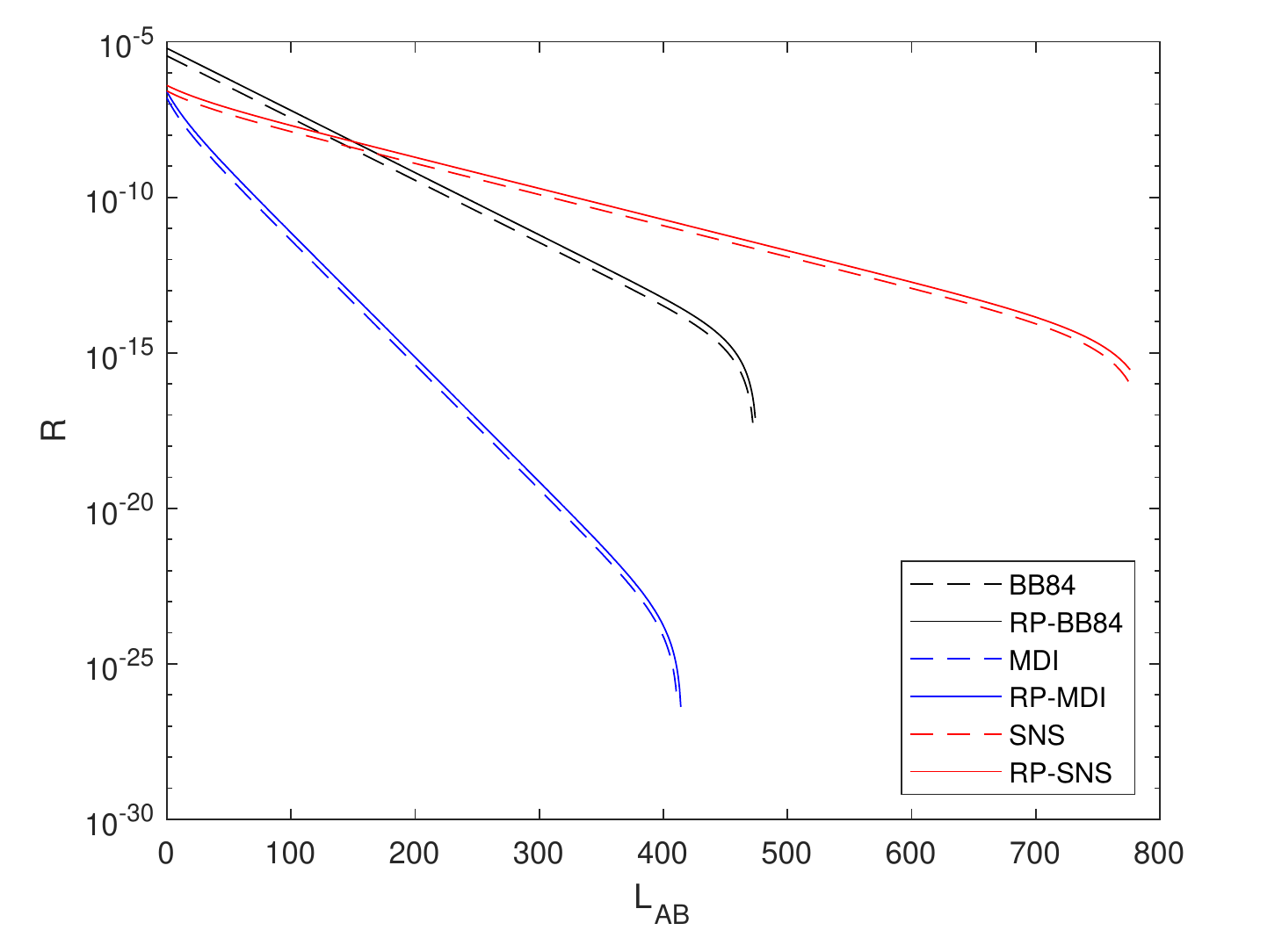}
	\caption{The signature rate with (solid lines) and without (dashed lines) random pairing in BB84-QDS (black), MDI-QDS (blue) and SNS-QDS (red) under misalignment error $e_d=5\%$. Other system parameters are shown in table~\ref{t_p}.  The signature rate with RP is higher than that witout RP by around $80\%-100\%$, in all protocols calculated here.}
	\label{pp}
\end{figure}

In this work, we propose a novel method of random pairing, which can be applied to any type of KGP in QDS protocols. In Fig~\ref{pp}, we show that the signature rate can be increased drastically by our random pairing method in BB84-QDS~\cite{amiri2016secure}, MDI-QDS~\cite{puthoor2016measurement} and SNS-QDS~\cite{WangTFQDS2021}. 

In reality, the data size is always finite. We study the RP-SNS-QDS with finite pluses through novel optimization. Numerical results show that the signature rates can be increased dramatically with random pairing and this advantage becomes more and more obvious with the increase of distance and misalignment error rate. 

{\bf{Acknowledgements:}} 
We thank Zong-Wen Yu and Xiao-Long Hu for helpful discussions. We acknowledge the financial support in part by Ministry of Science and Technology of China through The National Key Research and Development Program of China grant No. 2017YFA0303901; National Natural Science Foundation of China grant No.11774198 and No.11974204; Open Research Fund Program of the State Key
Laboratory of Low-Dimensional Quantum Physics No.KF202110.

\appendix

\section{Relevant parameters in RP-SNS-QDS} \label{a1}

In RP-SNS-QDS protocol, we need to know the bit-flip error rate $E$ and phase-flip error rate $e^{\mathrm{ph}}$ of the original bit string before random pairing. Based on \cite{wang2018twin}, in SNS-KGP, we can define four kinds of events in $Z$ windows

\begin{Def} ($C_0$ window, $C_1$ window, $\mathcal{B}$ window and $\mathcal{O}$ window in RP-SNS-QDS).
	A time window when  Alice decides to not send (send) the coherent state and Bob decides to send (not send) the coherent state is $C_0$ ($C_1$) window. A time window when both parties decide to send the coherent state or when both parties decide not to send the coherent state is 
	$\mathcal{B}$ window or $\mathcal{O}$ window.
\end{Def}

The counting rate $S_{\mathcal{K}}^d=\frac{n^d_{\mathcal{K}}}{N_{\mathcal{K}}}$ are as follows, where $N_{\mathcal{K}}$ is the number of $\mathcal{K}\in\{C_0, C_1,\mathcal{B},\mathcal{O}\}$ windows in $Z$ windows and $n^d_\mathcal{K}$ is the number of effective $\mathcal{K}$ windows heralded by detector $d$ ($d=l(r) $ represents the left (right) detector ) in
$Z$ windows,
\begin{equation}\label{}
	\begin{split}
		&S_{\mathcal{B}}^{l}=(1-p_d)e^{-\eta\mu}I_0(\eta\mu)-(1-p_d)^2e^{-2\eta\mu}=S_{\mathcal{B}}^{r},\\
		&S_{C_0}^{l}=(1-p_d)e^{-\eta \mu/2}-(1-p_d)^2e^{-\eta\mu}=S_{C_0}^{r},\\
		&S_{C_1}^{l}=S_{C_1}^{r}=S_{C_0}^{r},\\
		&S_{\mathcal{O}}^{l}=p_d(1-p_d)=S_{\mathcal{O}}^{r},\\
	\end{split}
\end{equation}
where $\eta=10^{-\frac{\alpha l_{0}}{10}}\eta_d$ represents the total efficiency, $l_0$ is the distance between Alice and Eve, $\eta_d$ is the detection efficiency, $p_d$ is the dark count rate of detectors, $\mu$ is the intensity of the coherent state and $I_0(x)$ is the $0$-order hyperbolic Bessel function of the first kind.

The bit-flip error rate $E$ of bit string before random pairing is 
\begin{equation}
	\begin{split}\label{SNS_E}
		&E = \frac{n_{\mathcal{B}}+n_{\mathcal{O}}}{{N_{t}}},\\
	\end{split}
\end{equation}
in which
\begin{equation}
	\begin{split}\label{}
		& n_{\mathcal{B}}=Np_z^2q^2(S_{\mathcal{B}}^{l}+S_{\mathcal{B}}^{r}),\\
	\end{split}
\end{equation}
\begin{equation}
	\begin{split}\label{}
		& n_{\mathcal{O}}=Np_z^2(1-q)^2(S_{\mathcal{O}}^{l}+S_{\mathcal{O}}^{r}),\\
	\end{split}
\end{equation}
\begin{equation}
	\begin{split}\label{}
		& n_{C_0}=Np_z^2q(1-q)(S_{C_0}^{l}+S_{C_0}^{r}),\\
	\end{split}
\end{equation}
\begin{equation}
	\begin{split}\label{}
		& n_{C_1}=Np_z^2q(1-q)(S_{C_1}^{l}+S_{C_1}^{r}),\\
	\end{split}
\end{equation}
\begin{equation}
	\begin{split}\label{}
		& N_t=n_{\mathcal{B}}+n_{\mathcal{O}}+n_{C_0}+n_{C_1},\\
	\end{split}
\end{equation}
where $N$ is the total number of pluses, $q$ is the probability of sending and $p_z$ the probability of choosing the signal window.

According to the definition of untagged bits, the proportion of untagged bits is
\begin{equation}
	\begin{split}\label{SNS_u}
		&\Delta_{\mathrm{un}}=\frac{n_1^0+n_1^1}{{N_{t}}},\\
	\end{split}
\end{equation}
\begin{equation}
	\begin{split}\label{}
		&n_1^0=Np_z^2q(1-q)\mu e^{-\mu}s_1=n_1^1,\\
	\end{split}
\end{equation}
where $n_1^0$ and $n_1^1$ represent the number of effective bits of single-photon state corresponding to $C_0$ and $C_1$ windows in $Z$ windows; $s_1$ is the counting rate of single-photon states.

For infinite decoy states, the counting rate of single-photon states and phase-flip error rate can be estimated as

\begin{equation}\label{}
	\begin{split}
		&s_{1} = (1-p_d)(\eta+2p_d(1-\eta)),\\
	\end{split}
\end{equation}

\begin{equation}
	\begin{split}
		&e^{\mathrm{ph},0} = p_d(1-p_d)(1-\eta)/s_{1},\\
	\end{split}
\end{equation}
and after considering the misalignment error $e_d$, it becomes
\begin{equation}\label{SNS_eph}
	\begin{split}
		&e^{\mathrm{ph}} = e_d(1-2e^{\mathrm{ph},0})+e^{\mathrm{ph},0}.\\
	\end{split}
\end{equation}

According to the above parameters ($E$, $\Delta_{\mathrm{un}}$, $e^{\mathrm{ph}} $) and the iteration relations in the main text, we can optimize the signature rate $R$ by choosing appropriate parameters given security level $\varepsilon$.

\section{Relevant parameters in RP-SCF-QDS}
We can obtain the bit-flip error rate $E$ and phase-flip error rate $e^{\mathrm{ph}}$ of the original bit string generated by SCF-KGP before random pairing according to \cite{wang2019practical}. 

In SCF-KGP, the counting rate $S_{\mathcal{V}}^d=\frac{n^d_{\mathcal{V}}}{N_{\mathcal{V}}}$ are as follows, where $N_{\mathcal{V}}$ is the number of $\mathcal{V}\in\{\tilde{Z},\mathcal{B},\mathcal{O}\}$ windows in set $v$ and $n^d_{\mathcal{V}}$ is the number of effective $\mathcal{V}$ windows heralded by detector $d$ ($d=l\ (r) $ represents the left (right) detector ) in set $v$,

\begin{equation}\label{Sad}
	\begin{split}
		&S_{\mathcal{B}}^{l}=e^{-2\eta\mu} p_d(1-p_d)+(1-e^{-2\eta\mu})(1-p_d),\\
		&S_{\mathcal{B}}^{r}=e^{-2\eta\mu} (1-p_d)p_d,\\
		&S_{\tilde{Z}}^{l}=(1-p_d)e^{-\eta \mu/2}-(1-p_d)^2e^{-\eta\mu}=S_{\tilde{Z}}^{r},\\
		&S_{\mathcal{O}}^{l}=p_d(1-p_d)=S_{\mathcal{O}}^{r},\\
	\end{split}
\end{equation}
where $\eta=10^{-\frac{\alpha l_{0}}{10}}\eta_d$ represents the total efficiency, $l_0$ is the distance between Alice and Eve, $\eta_d$ is the detection efficiency, $p_d$ is the dark count rate of detectors and $\mu$ is the intensity of the coherent state. 

After taking into account the misalignment error $e_d$, these counting rates become 

\begin{equation}\label{}
	\begin{split}
		&S_{\mathcal{V}}^{\prime d}=(1-e_d)S_{\mathcal{V}}^{d}+e_dS_{\mathcal{V}}^{d^{\prime}},\\
	\end{split}
\end{equation}
where $d\neq d^{\prime}$ and $d,d^{\prime}\in\{\mathcal{L},\mathcal{R}\}$.

The bit-flip error rate before random pairing is 
\begin{equation}\label{scf_Ev}
	\begin{split}
		&E=\frac{n^{v}_{\mathcal{B}}+n_{\mathcal{O}}^{v}}{n_v},\\
	\end{split}
\end{equation}
in which
\begin{equation}\label{}
	\begin{split}
		&n^{v}_{\mathcal{B}}=N\gamma_vq^2(S_{\mathcal{B}}^{\prime \mathcal{L}}+S_{\mathcal{B}}^{\prime \mathcal{R}}),\\
	\end{split}
\end{equation}
\begin{equation}\label{}
	\begin{split}
		&n^{v}_{\mathcal{O}}=N\gamma_v(1-q)^2(S_{\mathcal{O}}^{\prime \mathcal{L}}+S_{\mathcal{O}}^{\prime \mathcal{R}}),\\
	\end{split}
\end{equation}
\begin{equation}\label{}
	n_{\tilde{Z}}^v=2N\gamma_vq(1-q)(S_{\tilde{Z}}^{\prime\mathcal{L}}+S_{\tilde{Z}}^{\prime\mathcal{R}}),
\end{equation}
\begin{equation}\label{}
	n_{v}=n^{v}_{\mathcal{B}}+n^{v}_{\mathcal{O}}+n_{\tilde{Z}}^v,
\end{equation}
where $N$ is the total number of pluses, $q$ is the probability of sending, $\gamma_v=0.1$ in our simulation and $n_v$ is the number of effective events in set $v$.

The proportion of untagged bits before random pairing is

\begin{equation}\label{scf_u}
	\Delta_{\mathrm{un}} = \frac{n^v_{\tilde{Z}}}{n_v}.
\end{equation}

According to some observed data in set $v$, the upper bound of phase-flip error rate before random pairing is
\begin{equation}\label{scf_eph}
	e^{\mathrm{ph}}\le \overline{e}^{\mathrm{ph}}= \frac{(1+e^{-\mu})[\overline{S}^{\mathcal{R}}_{X_+}-\underline{S}^{\mathcal{L}}_{X_+}]+2S^{\prime L}_{\tilde{Z}}}{2(S_{\tilde{Z}}^{\prime {\mathcal{L}}}+S_{\tilde{Z}}^{\prime {\mathcal{R}}})},
\end{equation}
where $\overline{S}^{d}_{X_+}$ and $\underline{S}^{d}_{X_+}$ are the upper bound and lower bound of ${S}^{d}_{X_+}$ and ${S}^{d}_{X_+}$, respectively, $d\in\{\mathcal{L},\mathcal{R}\}$, and
\begin{equation}
	\begin{split}
		S^d_{X_+}\le \overline{S}^d_{X_+}=& \frac{1}{2(1+e^{-\mu})}\big\{e^{-\mu}S_{\mathcal{O}}^{\prime d}+\frac{1}{e^{-\mu}}S_{\mathcal{B}}^{\prime d}\\
		+&\frac{(1-e^{-\mu})^2}{e^{-\mu}}+2\sqrt{S_{\mathcal{O}}^{\prime d}S_{\mathcal{B}}^{\prime d}}\\
		+&2(1-e^{-\mu})\sqrt{S_{\mathcal{O}}^{\prime d}}
		+\frac{2(1-e^{-\mu})}{e^{-\mu}}\sqrt{S_{\mathcal{B}}^{\prime d}}\big\},\\
	\end{split}
\end{equation}
\begin{equation}
	\begin{split}
		S^d_{X_+}\ge \underline{S}^d_{X_+}=& \frac{1}{2(1+e^{-\mu})}\big\{e^{-\mu}S_{\mathcal{O}}^{\prime d}+\frac{1}{e^{-\mu}}S_{\mathcal{B}}^{\prime d}\\
		-&\big[2\sqrt{S_{\mathcal{O}}^{\prime d}S_{\mathcal{B}}^{\prime d}}+2(1-e^{-\mu})\sqrt{S_{\mathcal{O}}^{\prime d}}\\
		+&\frac{2(1-e^{-\mu})}{e^{-\mu}}\sqrt{S_{\mathcal{B}}^{\prime d}}\big]\big\}.\\
	\end{split}
\end{equation}

Using the above parameters ($E$, $\Delta_{\mathrm{un}}$, $e^{\mathrm{ph}}$) and the iteration relations in the main text, we can choose appropriate parameters to optimize the signature rate $R$ given security level $\varepsilon$.

\section{Relevant parameters in RP-SNS-QDS with finite-key effects} 

In the practical RP-SNS-KGP, the bit-flip error rate $E$ of bit string before random can be obtained from Eq.~\eqref{SNS_E} and then the core is how to obtain the counting rate of single-photon states $s_1$ and phase-flip error rate $e^{\mathrm{ph}}$ of original bit string before random pairing.  
According to \cite{Yu2019SNSdecoy,Jiang2019SNSfinite}, we use notations $\langle s_1\rangle$ and $\langle e^{\mathrm{ph}} \rangle$ to represent the expected values, then the corresponding lower bound and upper bound are
\begin{equation}
	\langle s_1\rangle \geq \langle s_{1}^{\mathrm{L}} \rangle=\frac{1}{2}(\langle s^{\mathrm{L}}_{01}\rangle+\langle s^{\mathrm{L}}_{10}\rangle),
\end{equation}

\begin{equation}
	\langle e^{\mathrm{ph}} \rangle\le \langle e^{\mathrm{ph},\mathrm{U}} \rangle=\frac{\langle T_{\Delta}^{\mathrm{U}}\rangle-\frac{1}{2}e^{-2\mu_1}\langle S^{\mathrm{L}}_{00}\rangle}{2\mu_1e^{-2\mu_1}\langle s_1^{\mathrm{L}}\rangle},
\end{equation}
where $\langle s^{\mathrm{L}}_{01}\rangle$, $\langle s^{\mathrm{L}}_{10}\rangle$ are shown in Eq.~\eqref{s01}; $\langle T_{\Delta}^{\mathrm{U}}\rangle$ is shown in Eq.~\eqref{TU}. In the decoy window, Alice (Bob) chooses phase-randomized coherent states with three different intensities $\mu_0=0$, $\mu_1$, $\mu_2$ and $S_{ij}$ $i,j\in\{0,1,2\}$, represents counting rate of source $ij$ when Alice (Bob) chooses phase-randomized coherent states with intensity $\mu_i$ ($\mu_j$).

We use Chernoff bound~\cite{chernoff} to estimate expected values $\phi$ from observed values, which are obtained from experiments directly,
\begin{equation}
	\begin{split}
		&\phi^{\mathrm{L}}(X)=\frac{X}{1+\delta_1(X)},\\
		&\phi^{\mathrm{U}}(X)=\frac{X}{1-\delta_2(X)},\\
	\end{split}
\end{equation}	
where $\delta_1(X)$ and $\delta_2(X)$ can be calculated from

\begin{equation}
	\begin{split}
		&(\frac{e^{\delta_1}}{(1+\delta_1)^{1+\delta_1}})^{\frac{X}{1+\delta_1}}=\frac{\xi}{2},\\
		&(\frac{e^{-\delta_2}}{(1-\delta_2)^{1-\delta_2}})^{\frac{X}{1-\delta_2}}=\frac{\xi}{2},\\
	\end{split}
\end{equation}	
where $\xi$ is the failure probability of estimation.

Using decoy-state method, we can obtain
\begin{equation}\label{s01}
	\begin{split}
		&\langle s_{01}^{\mathrm{L}}\rangle=\frac{\mu_2^2e^{\mu_1}\langle S_{01}^{\mathrm{L}}\rangle-\mu_1^2e^{\mu_2}\langle S_{02}^{\mathrm{U}}\rangle-(\mu_2^2-\mu_1^2)\langle S^{\mathrm{U}}_{00}\rangle}{\mu_1\mu_2(\mu_2-\mu_1)},\\
		&\langle s_{10}^{\mathrm{L}}\rangle=\frac{\mu_2^2e^{\mu_1}\langle S_{10}^{\mathrm{L}}\rangle-\mu_1^2e^{\mu_2}\langle S_{20}^{\mathrm{U}}\rangle-(\mu_2^2-\mu_1^2)\langle S^{\mathrm{U}}_{00}\rangle}{\mu_1\mu_2(\mu_2-\mu_1)},\\
	\end{split}
\end{equation}
Based on the analysis in~\cite{Yu2019SNSdecoy,Jiang2019SNSfinite}, notation $C_{\Delta^+}$ ($C_{\Delta^-}$) representing all instances where Alice and Bob prepare $\ket{e^{i\theta_A}\sqrt{\mu_1}}$ and $\ket{e^{i\theta_B}\sqrt{\mu_1}}$ when $|\delta_A-\delta_B|\le\Delta/2$ ($|\delta_A-\delta_B-\pi|\le\Delta/2$) is used to estimate $e^{\mathrm{ph},\mathrm{U}}$. There are $N_{\Delta^{\pm}}$ instances in $C_{\Delta^{\pm}}$
\begin{equation}
	N_{\Delta^{\pm}}=\frac{\Delta}{2\pi}(1-p_z)^2p_1^2N,
\end{equation}
We use $n_{\Delta^{+}}^{l}$ ($n_{\Delta^{-}}^{r}$) to represent effective events heralded by left (right) detector 
\begin{equation}
	n_{\Delta^{+}}^{l} =n_{\Delta^{-}}^{r}= [T_X(1-2e_d)+e_dS_X]N_{\Delta^{\pm}},
\end{equation}
where
\begin{equation}
	T_X=\frac{1}{\Delta}\int_{-\frac{\Delta}{2}}^{\frac{\Delta}{2}}(1-p_d)e^{-2\eta\mu_1\cos^2{\frac{\delta}{2}}}d\delta-(1-p_d)^2e^{-2\eta\mu_1},
\end{equation}

\begin{equation}
	\begin{split}
		S_X=&\frac{1}{\Delta}\int_{-\frac{\Delta}{2}}^{\frac{\Delta}{2}}(1-p_d)e^{-2\eta\mu_1\sin^2{\frac{\delta}{2}}}d\delta-(1-p_d)^2e^{-2\eta\mu_1}\\
		&+T_X.\\
	\end{split}
\end{equation}
Then, the counting error rate $\langle T_{\Delta}\rangle $ of $C_{\Delta^{\pm}}$ is
\begin{equation}\label{TU}
	\langle T_{\Delta}\rangle\le \langle T_{\Delta}^{\mathrm{U}} \rangle = \frac{\phi^{\mathrm{U}}(n_{\Delta^+}^{r}+n_{\Delta^-}^{l})}{2N_{\Delta^{\pm}}}.
\end{equation}
Next, we use Chernoff bound to estimate real values $\varphi$ of the specific experiment from expected values, 
\begin{equation}
	\begin{split}
		&\varphi^{\mathrm{L}}(Y)=[1+\delta_1^{\prime}(Y)]Y,\\
		&\varphi^{\mathrm{U}}(Y)=[1-\delta_2^{\prime}(Y)]Y,\\
	\end{split}
\end{equation}	
where $\delta_1^{\prime}(X)$ and $\delta_2^{\prime}(X)$ can be obtained from
\begin{equation}
	\begin{split}
		&(\frac{e^{\delta_1^{\prime}}}{(1+\delta_1^{\prime})^{1+\delta_1^{\prime}}})^{Y}=\frac{\xi}{2},\\
		&(\frac{e^{-\delta_2^{\prime}}}{(1-\delta_2^{\prime})^{1-\delta_2^{\prime}}})^{Y}=\frac{\xi}{2},\\
	\end{split}
\end{equation}	
where $\xi$ is the failure probability of estimation. 

Therefore, the proportion $\Delta_{u,f}$ of untagged bits of the original bit string is  
\begin{equation}
	\begin{split}\label{}
		&\Delta_{u,f}=\frac{\varphi^{\mathrm{L}}(\langle n_{1}^{\mathrm{L}}\rangle)}{{N_{t}}},\\
	\end{split}
\end{equation}
where
\begin{equation}
	\langle n_{1}^{\mathrm{L}}\rangle=2Np_z^2q(1-q)\mu e^{-\mu} \langle s_1^{\mathrm{L}}\rangle,
\end{equation}
and
\begin{equation}\label{feph}
	e^{\mathrm{ph}}=\frac{\varphi^{\mathrm{U}}(N_{t}\Delta_{u,f}\langle e^{\mathrm{ph},\mathrm{U}}\rangle )}{N_{t}\Delta_{u,f}}.
\end{equation}

The lower bound (upper bound) of the number $n_x$  of bit pairs consisting of $x$ errors, $x\in\{0,1,2\}$, can be calculated by

\begin{equation}
	\begin{split}\label{}
		&n_x^{\mathrm{L}}=\varphi^{\mathrm{L}}(n_x),\\
		&n_x^{\mathrm{R}}=\varphi^{\mathrm{R}}(n_x),\\
	\end{split}
\end{equation}
where $n_x$ is shown in main text.
Similarly, the lower bound of the number of bit pairs consisted of one tagged bit and one untagged bit is
\begin{equation}
	\begin{split}\label{fen}
		&n_{\mathrm{ut}}^{\mathrm{L}}=\varphi^{\mathrm{L}}(n_{\mathrm{ut}}),\\
	\end{split}
\end{equation}
where $n_{\mathrm{ut}}$ is shown in main text. 
Based on $n_0^{\mathrm{L}}$, $n_2^{\mathrm{U}}$, $n_{\mathrm{ut}}^{\mathrm{L}}$ and $e^{\mathrm{ph}}$, we can obtain $\mathcal{H}_f$ in the main text.
Given system parameters $(\alpha, \eta_d,P_d,\varepsilon)$, we need to optimize the signature rate $R_f$  by choosing appropriate parameters $(\mu,q,\mu_1,\mu_2,p_z,p_0,\Delta,\gamma_T,\varepsilon_{PE},g,\xi)$.

\bibliography{refs}

\begin{thebibliography}{42}%
\makeatletter
\providecommand \@ifxundefined [1]{%
 \@ifx{#1\undefined}
}%
\providecommand \@ifnum [1]{%
 \ifnum #1\expandafter \@firstoftwo
 \else \expandafter \@secondoftwo
 \fi
}%
\providecommand \@ifx [1]{%
 \ifx #1\expandafter \@firstoftwo
 \else \expandafter \@secondoftwo
 \fi
}%
\providecommand \natexlab [1]{#1}%
\providecommand \enquote  [1]{``#1''}%
\providecommand \bibnamefont  [1]{#1}%
\providecommand \bibfnamefont [1]{#1}%
\providecommand \citenamefont [1]{#1}%
\providecommand \href@noop [0]{\@secondoftwo}%
\providecommand \href [0]{\begingroup \@sanitize@url \@href}%
\providecommand \@href[1]{\@@startlink{#1}\@@href}%
\providecommand \@@href[1]{\endgroup#1\@@endlink}%
\providecommand \@sanitize@url [0]{\catcode `\\12\catcode `\$12\catcode
  `\&12\catcode `\#12\catcode `\^12\catcode `\_12\catcode `\%12\relax}%
\providecommand \@@startlink[1]{}%
\providecommand \@@endlink[0]{}%
\providecommand \url  [0]{\begingroup\@sanitize@url \@url }%
\providecommand \@url [1]{\endgroup\@href {#1}{\urlprefix }}%
\providecommand \urlprefix  [0]{URL }%
\providecommand \Eprint [0]{\href }%
\providecommand \doibase [0]{http://dx.doi.org/}%
\providecommand \selectlanguage [0]{\@gobble}%
\providecommand \bibinfo  [0]{\@secondoftwo}%
\providecommand \bibfield  [0]{\@secondoftwo}%
\providecommand \translation [1]{[#1]}%
\providecommand \BibitemOpen [0]{}%
\providecommand \bibitemStop [0]{}%
\providecommand \bibitemNoStop [0]{.\EOS\space}%
\providecommand \EOS [0]{\spacefactor3000\relax}%
\providecommand \BibitemShut  [1]{\csname bibitem#1\endcsname}%
\let\auto@bib@innerbib\@empty
\bibitem [{\citenamefont {Diffie}\ and\ \citenamefont
  {Hellman}(1976)}]{1055638}%
  \BibitemOpen
  \bibfield  {author} {\bibinfo {author} {\bibfnamefont {W.}~\bibnamefont
  {Diffie}}\ and\ \bibinfo {author} {\bibfnamefont {M.}~\bibnamefont
  {Hellman}},\ }\bibfield  {title} {\enquote {\bibinfo {title} {New directions
  in cryptography},}\ }\href@noop {} {\bibfield  {journal} {\bibinfo  {journal}
  {IEEE Transactions on Information Theory}\ }\textbf {\bibinfo {volume}
  {22}},\ \bibinfo {pages} {644--654} (\bibinfo {year} {1976})}\BibitemShut
  {NoStop}%
\bibitem [{\citenamefont {Gottesman}\ and\ \citenamefont
  {Chuang}(2001)}]{gottesman2001quantum}%
  \BibitemOpen
  \bibfield  {author} {\bibinfo {author} {\bibfnamefont {Daniel}\ \bibnamefont
  {Gottesman}}\ and\ \bibinfo {author} {\bibfnamefont {Isaac}\ \bibnamefont
  {Chuang}},\ }\bibfield  {title} {\enquote {\bibinfo {title} {Quantum digital
  signatures},}\ }\href@noop {} {\bibfield  {journal} {\bibinfo  {journal}
  {arXiv preprint quant-ph/0105032}\ } (\bibinfo {year} {2001})}\BibitemShut
  {NoStop}%
\bibitem [{\citenamefont {Andersson}\ \emph {et~al.}(2006)\citenamefont
  {Andersson}, \citenamefont {Curty},\ and\ \citenamefont
  {Jex}}]{AnderssonQDS2006}%
  \BibitemOpen
  \bibfield  {author} {\bibinfo {author} {\bibfnamefont {Erika}\ \bibnamefont
  {Andersson}}, \bibinfo {author} {\bibfnamefont {Marcos}\ \bibnamefont
  {Curty}}, \ and\ \bibinfo {author} {\bibfnamefont {Igor}\ \bibnamefont
  {Jex}},\ }\bibfield  {title} {\enquote {\bibinfo {title} {Experimentally
  realizable quantum comparison of coherent states and its applications},}\
  }\href@noop {} {\bibfield  {journal} {\bibinfo  {journal} {Phys. Rev. A}\
  }\textbf {\bibinfo {volume} {74}},\ \bibinfo {pages} {022304} (\bibinfo
  {year} {2006})}\BibitemShut {NoStop}%
\bibitem [{\citenamefont {Dunjko}\ \emph {et~al.}(2014)\citenamefont {Dunjko},
  \citenamefont {Wallden},\ and\ \citenamefont
  {Andersson}}]{dunjko2014quantum}%
  \BibitemOpen
  \bibfield  {author} {\bibinfo {author} {\bibfnamefont {Vedran}\ \bibnamefont
  {Dunjko}}, \bibinfo {author} {\bibfnamefont {Petros}\ \bibnamefont
  {Wallden}}, \ and\ \bibinfo {author} {\bibfnamefont {Erika}\ \bibnamefont
  {Andersson}},\ }\bibfield  {title} {\enquote {\bibinfo {title} {Quantum
  digital signatures without quantum memory},}\ }\href@noop {} {\bibfield
  {journal} {\bibinfo  {journal} {Phys. Rev. Lett.}\ }\textbf {\bibinfo
  {volume} {112}},\ \bibinfo {pages} {040502} (\bibinfo {year}
  {2014})}\BibitemShut {NoStop}%
\bibitem [{\citenamefont {Amiri}\ \emph {et~al.}(2016)\citenamefont {Amiri},
  \citenamefont {Wallden}, \citenamefont {Kent},\ and\ \citenamefont
  {Andersson}}]{amiri2016secure}%
  \BibitemOpen
  \bibfield  {author} {\bibinfo {author} {\bibfnamefont {Ryan}\ \bibnamefont
  {Amiri}}, \bibinfo {author} {\bibfnamefont {Petros}\ \bibnamefont {Wallden}},
  \bibinfo {author} {\bibfnamefont {Adrian}\ \bibnamefont {Kent}}, \ and\
  \bibinfo {author} {\bibfnamefont {Erika}\ \bibnamefont {Andersson}},\
  }\bibfield  {title} {\enquote {\bibinfo {title} {Secure quantum signatures
  using insecure quantum channels},}\ }\href@noop {} {\bibfield  {journal}
  {\bibinfo  {journal} {Phys. Rev. A}\ }\textbf {\bibinfo {volume} {93}},\
  \bibinfo {pages} {032325} (\bibinfo {year} {2016})}\BibitemShut {NoStop}%
\bibitem [{\citenamefont {Yin}\ \emph {et~al.}(2016{\natexlab{a}})\citenamefont
  {Yin}, \citenamefont {Fu},\ and\ \citenamefont {Chen}}]{yin2016practical}%
  \BibitemOpen
  \bibfield  {author} {\bibinfo {author} {\bibfnamefont {Hua-Lei}\ \bibnamefont
  {Yin}}, \bibinfo {author} {\bibfnamefont {Yao}\ \bibnamefont {Fu}}, \ and\
  \bibinfo {author} {\bibfnamefont {Zeng-Bing}\ \bibnamefont {Chen}},\
  }\bibfield  {title} {\enquote {\bibinfo {title} {Practical quantum digital
  signature},}\ }\href@noop {} {\bibfield  {journal} {\bibinfo  {journal}
  {Phys. Rev. A}\ }\textbf {\bibinfo {volume} {93}},\ \bibinfo {pages} {032316}
  (\bibinfo {year} {2016}{\natexlab{a}})}\BibitemShut {NoStop}%
\bibitem [{\citenamefont {Collins}\ \emph {et~al.}(2014)\citenamefont
  {Collins}, \citenamefont {Donaldson}, \citenamefont {Dunjko}, \citenamefont
  {Wallden}, \citenamefont {Clarke}, \citenamefont {Andersson}, \citenamefont
  {Jeffers},\ and\ \citenamefont {Buller}}]{collins2014realization}%
  \BibitemOpen
  \bibfield  {author} {\bibinfo {author} {\bibfnamefont {Robert~J.}\
  \bibnamefont {Collins}}, \bibinfo {author} {\bibfnamefont {Ross~J.}\
  \bibnamefont {Donaldson}}, \bibinfo {author} {\bibfnamefont {Vedran}\
  \bibnamefont {Dunjko}}, \bibinfo {author} {\bibfnamefont {Petros}\
  \bibnamefont {Wallden}}, \bibinfo {author} {\bibfnamefont {Patrick~J.}\
  \bibnamefont {Clarke}}, \bibinfo {author} {\bibfnamefont {Erika}\
  \bibnamefont {Andersson}}, \bibinfo {author} {\bibfnamefont {John}\
  \bibnamefont {Jeffers}}, \ and\ \bibinfo {author} {\bibfnamefont {Gerald~S.}\
  \bibnamefont {Buller}},\ }\bibfield  {title} {\enquote {\bibinfo {title}
  {Realization of quantum digital signatures without the requirement of quantum
  memory},}\ }\href@noop {} {\bibfield  {journal} {\bibinfo  {journal} {Phys.
  Rev. Lett.}\ }\textbf {\bibinfo {volume} {113}},\ \bibinfo {pages} {040502}
  (\bibinfo {year} {2014})}\BibitemShut {NoStop}%
\bibitem [{\citenamefont {Donaldson}\ \emph {et~al.}(2016)\citenamefont
  {Donaldson}, \citenamefont {Collins}, \citenamefont {Kleczkowska},
  \citenamefont {Amiri}, \citenamefont {Wallden}, \citenamefont {Dunjko},
  \citenamefont {Jeffers}, \citenamefont {Andersson},\ and\ \citenamefont
  {Buller}}]{donaldson2016experimental}%
  \BibitemOpen
  \bibfield  {author} {\bibinfo {author} {\bibfnamefont {Ross~J.}\ \bibnamefont
  {Donaldson}}, \bibinfo {author} {\bibfnamefont {Robert~J.}\ \bibnamefont
  {Collins}}, \bibinfo {author} {\bibfnamefont {Klaudia}\ \bibnamefont
  {Kleczkowska}}, \bibinfo {author} {\bibfnamefont {Ryan}\ \bibnamefont
  {Amiri}}, \bibinfo {author} {\bibfnamefont {Petros}\ \bibnamefont {Wallden}},
  \bibinfo {author} {\bibfnamefont {Vedran}\ \bibnamefont {Dunjko}}, \bibinfo
  {author} {\bibfnamefont {John}\ \bibnamefont {Jeffers}}, \bibinfo {author}
  {\bibfnamefont {Erika}\ \bibnamefont {Andersson}}, \ and\ \bibinfo {author}
  {\bibfnamefont {Gerald~S.}\ \bibnamefont {Buller}},\ }\bibfield  {title}
  {\enquote {\bibinfo {title} {Experimental demonstration of kilometer-range
  quantum digital signatures},}\ }\href@noop {} {\bibfield  {journal} {\bibinfo
   {journal} {Phys. Rev. A}\ }\textbf {\bibinfo {volume} {93}},\ \bibinfo
  {pages} {012329} (\bibinfo {year} {2016})}\BibitemShut {NoStop}%
\bibitem [{\citenamefont {Croal}\ \emph {et~al.}(2016)\citenamefont {Croal},
  \citenamefont {Peuntinger}, \citenamefont {Heim}, \citenamefont {Khan},
  \citenamefont {Marquardt}, \citenamefont {Leuchs}, \citenamefont {Wallden},
  \citenamefont {Andersson},\ and\ \citenamefont {Korolkova}}]{croal2016free}%
  \BibitemOpen
  \bibfield  {author} {\bibinfo {author} {\bibfnamefont {Callum}\ \bibnamefont
  {Croal}}, \bibinfo {author} {\bibfnamefont {Christian}\ \bibnamefont
  {Peuntinger}}, \bibinfo {author} {\bibfnamefont {Bettina}\ \bibnamefont
  {Heim}}, \bibinfo {author} {\bibfnamefont {Imran}\ \bibnamefont {Khan}},
  \bibinfo {author} {\bibfnamefont {Christoph}\ \bibnamefont {Marquardt}},
  \bibinfo {author} {\bibfnamefont {Gerd}\ \bibnamefont {Leuchs}}, \bibinfo
  {author} {\bibfnamefont {Petros}\ \bibnamefont {Wallden}}, \bibinfo {author}
  {\bibfnamefont {Erika}\ \bibnamefont {Andersson}}, \ and\ \bibinfo {author}
  {\bibfnamefont {Natalia}\ \bibnamefont {Korolkova}},\ }\bibfield  {title}
  {\enquote {\bibinfo {title} {Free-space quantum signatures using heterodyne
  measurements},}\ }\href@noop {} {\bibfield  {journal} {\bibinfo  {journal}
  {Phys. Rev. Lett.}\ }\textbf {\bibinfo {volume} {117}},\ \bibinfo {pages}
  {100503} (\bibinfo {year} {2016})}\BibitemShut {NoStop}%
\bibitem [{\citenamefont {Yin}\ \emph {et~al.}(2017)\citenamefont {Yin},
  \citenamefont {Fu}, \citenamefont {Liu}, \citenamefont {Tang}, \citenamefont
  {Wang}, \citenamefont {You}, \citenamefont {Zhang}, \citenamefont {Chen},
  \citenamefont {Wang}, \citenamefont {Zhang}, \citenamefont {Chen},
  \citenamefont {Chen},\ and\ \citenamefont {Pan}}]{yin2017experimental}%
  \BibitemOpen
  \bibfield  {author} {\bibinfo {author} {\bibfnamefont {Hua-Lei}\ \bibnamefont
  {Yin}}, \bibinfo {author} {\bibfnamefont {Yao}\ \bibnamefont {Fu}}, \bibinfo
  {author} {\bibfnamefont {Hui}\ \bibnamefont {Liu}}, \bibinfo {author}
  {\bibfnamefont {Qi-Jie}\ \bibnamefont {Tang}}, \bibinfo {author}
  {\bibfnamefont {Jian}\ \bibnamefont {Wang}}, \bibinfo {author} {\bibfnamefont
  {Li-Xing}\ \bibnamefont {You}}, \bibinfo {author} {\bibfnamefont {Wei-Jun}\
  \bibnamefont {Zhang}}, \bibinfo {author} {\bibfnamefont {Si-Jing}\
  \bibnamefont {Chen}}, \bibinfo {author} {\bibfnamefont {Zhen}\ \bibnamefont
  {Wang}}, \bibinfo {author} {\bibfnamefont {Qiang}\ \bibnamefont {Zhang}},
  \bibinfo {author} {\bibfnamefont {Teng-Yun}\ \bibnamefont {Chen}}, \bibinfo
  {author} {\bibfnamefont {Zeng-Bing}\ \bibnamefont {Chen}}, \ and\ \bibinfo
  {author} {\bibfnamefont {Jian-Wei}\ \bibnamefont {Pan}},\ }\bibfield  {title}
  {\enquote {\bibinfo {title} {Experimental quantum digital signature over 102
  km},}\ }\href@noop {} {\bibfield  {journal} {\bibinfo  {journal} {Phys. Rev.
  A}\ }\textbf {\bibinfo {volume} {95}},\ \bibinfo {pages} {032334} (\bibinfo
  {year} {2017})}\BibitemShut {NoStop}%
\bibitem [{\citenamefont {Roberts}\ \emph {et~al.}(2017)\citenamefont
  {Roberts}, \citenamefont {Lucamarini}, \citenamefont {Yuan}, \citenamefont
  {Dynes}, \citenamefont {Comandar}, \citenamefont {Sharpe}, \citenamefont
  {Shields}, \citenamefont {Curty}, \citenamefont {Puthoor},\ and\
  \citenamefont {Andersson}}]{roberts2017experimental}%
  \BibitemOpen
  \bibfield  {author} {\bibinfo {author} {\bibfnamefont {GL}~\bibnamefont
  {Roberts}}, \bibinfo {author} {\bibfnamefont {M}~\bibnamefont {Lucamarini}},
  \bibinfo {author} {\bibfnamefont {ZL}~\bibnamefont {Yuan}}, \bibinfo {author}
  {\bibfnamefont {JF}~\bibnamefont {Dynes}}, \bibinfo {author} {\bibfnamefont
  {LC}~\bibnamefont {Comandar}}, \bibinfo {author} {\bibfnamefont
  {AW}~\bibnamefont {Sharpe}}, \bibinfo {author} {\bibfnamefont
  {AJ}~\bibnamefont {Shields}}, \bibinfo {author} {\bibfnamefont
  {M}~\bibnamefont {Curty}}, \bibinfo {author} {\bibfnamefont {IV}~\bibnamefont
  {Puthoor}}, \ and\ \bibinfo {author} {\bibfnamefont {E}~\bibnamefont
  {Andersson}},\ }\bibfield  {title} {\enquote {\bibinfo {title} {Experimental
  measurement-device-independent quantum digital signatures},}\ }\href@noop {}
  {\bibfield  {journal} {\bibinfo  {journal} {Nature communications}\ }\textbf
  {\bibinfo {volume} {8}},\ \bibinfo {pages} {1--7} (\bibinfo {year}
  {2017})}\BibitemShut {NoStop}%
\bibitem [{\citenamefont {Zhang}\ \emph {et~al.}(2018)\citenamefont {Zhang},
  \citenamefont {Zhou}, \citenamefont {Ding}, \citenamefont {Zhang},
  \citenamefont {Guo},\ and\ \citenamefont {Wang}}]{zhang2018proof}%
  \BibitemOpen
  \bibfield  {author} {\bibinfo {author} {\bibfnamefont {Chun-Hui}\
  \bibnamefont {Zhang}}, \bibinfo {author} {\bibfnamefont {Xing-Yu}\
  \bibnamefont {Zhou}}, \bibinfo {author} {\bibfnamefont {Hua-Jian}\
  \bibnamefont {Ding}}, \bibinfo {author} {\bibfnamefont {Chun-Mei}\
  \bibnamefont {Zhang}}, \bibinfo {author} {\bibfnamefont {Guang-Can}\
  \bibnamefont {Guo}}, \ and\ \bibinfo {author} {\bibfnamefont {Qin}\
  \bibnamefont {Wang}},\ }\bibfield  {title} {\enquote {\bibinfo {title}
  {Proof-of-principle demonstration of passive decoy-state quantum digital
  signatures over 200 km},}\ }\href@noop {} {\bibfield  {journal} {\bibinfo
  {journal} {Physical Review Applied}\ }\textbf {\bibinfo {volume} {10}},\
  \bibinfo {pages} {034033} (\bibinfo {year} {2018})}\BibitemShut {NoStop}%
\bibitem [{\citenamefont {Ding}\ \emph {et~al.}(2020)\citenamefont {Ding},
  \citenamefont {Chen}, \citenamefont {Ji}, \citenamefont {Zhou}, \citenamefont
  {Zhang}, \citenamefont {Zhang},\ and\ \citenamefont {Wang}}]{ding2020280}%
  \BibitemOpen
  \bibfield  {author} {\bibinfo {author} {\bibfnamefont {Hua-Jian}\
  \bibnamefont {Ding}}, \bibinfo {author} {\bibfnamefont {Jing-Jing}\
  \bibnamefont {Chen}}, \bibinfo {author} {\bibfnamefont {Liang}\ \bibnamefont
  {Ji}}, \bibinfo {author} {\bibfnamefont {Xing-Yu}\ \bibnamefont {Zhou}},
  \bibinfo {author} {\bibfnamefont {Chun-Hui}\ \bibnamefont {Zhang}}, \bibinfo
  {author} {\bibfnamefont {Chun-Mei}\ \bibnamefont {Zhang}}, \ and\ \bibinfo
  {author} {\bibfnamefont {Qin}\ \bibnamefont {Wang}},\ }\bibfield  {title}
  {\enquote {\bibinfo {title} {280-km experimental demonstration of a quantum
  digital signature with one decoy state},}\ }\href@noop {} {\bibfield
  {journal} {\bibinfo  {journal} {Optics letters}\ }\textbf {\bibinfo {volume}
  {45}},\ \bibinfo {pages} {1711--1714} (\bibinfo {year} {2020})}\BibitemShut
  {NoStop}%
\bibitem [{\citenamefont {Pirandola}\ \emph {et~al.}(2020)\citenamefont
  {Pirandola}, \citenamefont {Andersen}, \citenamefont {Banchi}, \citenamefont
  {Berta}, \citenamefont {Bunandar}, \citenamefont {Colbeck}, \citenamefont
  {Englund}, \citenamefont {Gehring}, \citenamefont {Lupo}, \citenamefont
  {Ottaviani} \emph {et~al.}}]{pirandola2020advances}%
  \BibitemOpen
  \bibfield  {author} {\bibinfo {author} {\bibfnamefont {Stefano}\ \bibnamefont
  {Pirandola}}, \bibinfo {author} {\bibfnamefont {Ulrik~L}\ \bibnamefont
  {Andersen}}, \bibinfo {author} {\bibfnamefont {Leonardo}\ \bibnamefont
  {Banchi}}, \bibinfo {author} {\bibfnamefont {Mario}\ \bibnamefont {Berta}},
  \bibinfo {author} {\bibfnamefont {Darius}\ \bibnamefont {Bunandar}}, \bibinfo
  {author} {\bibfnamefont {Roger}\ \bibnamefont {Colbeck}}, \bibinfo {author}
  {\bibfnamefont {Dirk}\ \bibnamefont {Englund}}, \bibinfo {author}
  {\bibfnamefont {Tobias}\ \bibnamefont {Gehring}}, \bibinfo {author}
  {\bibfnamefont {Cosmo}\ \bibnamefont {Lupo}}, \bibinfo {author}
  {\bibfnamefont {Carlo}\ \bibnamefont {Ottaviani}},  \emph {et~al.},\
  }\bibfield  {title} {\enquote {\bibinfo {title} {Advances in quantum
  cryptography},}\ }\href@noop {} {\bibfield  {journal} {\bibinfo  {journal}
  {Advances in Optics and Photonics}\ }\textbf {\bibinfo {volume} {12}},\
  \bibinfo {pages} {1012--1236} (\bibinfo {year} {2020})}\BibitemShut {NoStop}%
\bibitem [{\citenamefont {Wallden}\ \emph {et~al.}(2015)\citenamefont
  {Wallden}, \citenamefont {Dunjko}, \citenamefont {Kent},\ and\ \citenamefont
  {Andersson}}]{wallden2015quantum}%
  \BibitemOpen
  \bibfield  {author} {\bibinfo {author} {\bibfnamefont {Petros}\ \bibnamefont
  {Wallden}}, \bibinfo {author} {\bibfnamefont {Vedran}\ \bibnamefont
  {Dunjko}}, \bibinfo {author} {\bibfnamefont {Adrian}\ \bibnamefont {Kent}}, \
  and\ \bibinfo {author} {\bibfnamefont {Erika}\ \bibnamefont {Andersson}},\
  }\bibfield  {title} {\enquote {\bibinfo {title} {Quantum digital signatures
  with quantum-key-distribution components},}\ }\href@noop {} {\bibfield
  {journal} {\bibinfo  {journal} {Phys. Rev. A}\ }\textbf {\bibinfo {volume}
  {91}},\ \bibinfo {pages} {042304} (\bibinfo {year} {2015})}\BibitemShut
  {NoStop}%
\bibitem [{\citenamefont {Puthoor}\ \emph {et~al.}(2016)\citenamefont
  {Puthoor}, \citenamefont {Amiri}, \citenamefont {Wallden}, \citenamefont
  {Curty},\ and\ \citenamefont {Andersson}}]{puthoor2016measurement}%
  \BibitemOpen
  \bibfield  {author} {\bibinfo {author} {\bibfnamefont {Ittoop~Vergheese}\
  \bibnamefont {Puthoor}}, \bibinfo {author} {\bibfnamefont {Ryan}\
  \bibnamefont {Amiri}}, \bibinfo {author} {\bibfnamefont {Petros}\
  \bibnamefont {Wallden}}, \bibinfo {author} {\bibfnamefont {Marcos}\
  \bibnamefont {Curty}}, \ and\ \bibinfo {author} {\bibfnamefont {Erika}\
  \bibnamefont {Andersson}},\ }\bibfield  {title} {\enquote {\bibinfo {title}
  {Measurement-device-independent quantum digital signatures},}\ }\href@noop {}
  {\bibfield  {journal} {\bibinfo  {journal} {Phys. Rev. A}\ }\textbf {\bibinfo
  {volume} {94}},\ \bibinfo {pages} {022328} (\bibinfo {year}
  {2016})}\BibitemShut {NoStop}%
\bibitem [{\citenamefont {Bennett}\ and\ \citenamefont
  {Brassard}(2014)}]{BENNETT20147}%
  \BibitemOpen
  \bibfield  {author} {\bibinfo {author} {\bibfnamefont {Charles~H.}\
  \bibnamefont {Bennett}}\ and\ \bibinfo {author} {\bibfnamefont {Gilles}\
  \bibnamefont {Brassard}},\ }\bibfield  {title} {\enquote {\bibinfo {title}
  {Quantum cryptography: Public key distribution and coin tossing},}\
  }\href@noop {} {\bibfield  {journal} {\bibinfo  {journal} {Theoretical
  Computer Science}\ }\textbf {\bibinfo {volume} {560}},\ \bibinfo {pages}
  {7--11} (\bibinfo {year} {2014})},\ \bibinfo {note} {theoretical Aspects of
  Quantum Cryptography – celebrating 30 years of BB84}\BibitemShut {NoStop}%
\bibitem [{\citenamefont {Hwang}(2003)}]{Hwang2003}%
  \BibitemOpen
  \bibfield  {author} {\bibinfo {author} {\bibfnamefont {Won-Young}\
  \bibnamefont {Hwang}},\ }\bibfield  {title} {\enquote {\bibinfo {title}
  {Quantum key distribution with high loss: Toward global secure
  communication},}\ }\href@noop {} {\bibfield  {journal} {\bibinfo  {journal}
  {Phys. Rev. Lett.}\ }\textbf {\bibinfo {volume} {91}},\ \bibinfo {pages}
  {057901} (\bibinfo {year} {2003})}\BibitemShut {NoStop}%
\bibitem [{\citenamefont {Wang}(2005)}]{Wangdecoy2005}%
  \BibitemOpen
  \bibfield  {author} {\bibinfo {author} {\bibfnamefont {Xiang-Bin}\
  \bibnamefont {Wang}},\ }\bibfield  {title} {\enquote {\bibinfo {title}
  {Beating the photon-number-splitting attack in practical quantum
  cryptography},}\ }\href@noop {} {\bibfield  {journal} {\bibinfo  {journal}
  {Phys. Rev. Lett.}\ }\textbf {\bibinfo {volume} {94}},\ \bibinfo {pages}
  {230503} (\bibinfo {year} {2005})}\BibitemShut {NoStop}%
\bibitem [{\citenamefont {Lo}\ \emph {et~al.}(2005)\citenamefont {Lo},
  \citenamefont {Ma},\ and\ \citenamefont {Chen}}]{Lodecoy2005}%
  \BibitemOpen
  \bibfield  {author} {\bibinfo {author} {\bibfnamefont {Hoi-Kwong}\
  \bibnamefont {Lo}}, \bibinfo {author} {\bibfnamefont {Xiongfeng}\
  \bibnamefont {Ma}}, \ and\ \bibinfo {author} {\bibfnamefont {Kai}\
  \bibnamefont {Chen}},\ }\bibfield  {title} {\enquote {\bibinfo {title} {Decoy
  state quantum key distribution},}\ }\href@noop {} {\bibfield  {journal}
  {\bibinfo  {journal} {Phys. Rev. Lett.}\ }\textbf {\bibinfo {volume} {94}},\
  \bibinfo {pages} {230504} (\bibinfo {year} {2005})}\BibitemShut {NoStop}%
\bibitem [{\citenamefont {Boaron}\ \emph {et~al.}(2018)\citenamefont {Boaron},
  \citenamefont {Boso}, \citenamefont {Rusca}, \citenamefont {Vulliez},
  \citenamefont {Autebert}, \citenamefont {Caloz}, \citenamefont {Perrenoud},
  \citenamefont {Gras}, \citenamefont {Bussi\`eres}, \citenamefont {Li},
  \citenamefont {Nolan}, \citenamefont {Martin},\ and\ \citenamefont
  {Zbinden}}]{PhysRevLett.121.190502}%
  \BibitemOpen
  \bibfield  {author} {\bibinfo {author} {\bibfnamefont {Alberto}\ \bibnamefont
  {Boaron}}, \bibinfo {author} {\bibfnamefont {Gianluca}\ \bibnamefont {Boso}},
  \bibinfo {author} {\bibfnamefont {Davide}\ \bibnamefont {Rusca}}, \bibinfo
  {author} {\bibfnamefont {C\'edric}\ \bibnamefont {Vulliez}}, \bibinfo
  {author} {\bibfnamefont {Claire}\ \bibnamefont {Autebert}}, \bibinfo {author}
  {\bibfnamefont {Misael}\ \bibnamefont {Caloz}}, \bibinfo {author}
  {\bibfnamefont {Matthieu}\ \bibnamefont {Perrenoud}}, \bibinfo {author}
  {\bibfnamefont {Ga\"etan}\ \bibnamefont {Gras}}, \bibinfo {author}
  {\bibfnamefont {F\'elix}\ \bibnamefont {Bussi\`eres}}, \bibinfo {author}
  {\bibfnamefont {Ming-Jun}\ \bibnamefont {Li}}, \bibinfo {author}
  {\bibfnamefont {Daniel}\ \bibnamefont {Nolan}}, \bibinfo {author}
  {\bibfnamefont {Anthony}\ \bibnamefont {Martin}}, \ and\ \bibinfo {author}
  {\bibfnamefont {Hugo}\ \bibnamefont {Zbinden}},\ }\bibfield  {title}
  {\enquote {\bibinfo {title} {Secure quantum key distribution over 421 km of
  optical fiber},}\ }\href@noop {} {\bibfield  {journal} {\bibinfo  {journal}
  {Phys. Rev. Lett.}\ }\textbf {\bibinfo {volume} {121}},\ \bibinfo {pages}
  {190502} (\bibinfo {year} {2018})}\BibitemShut {NoStop}%
\bibitem [{\citenamefont {Lo}\ \emph {et~al.}(2012)\citenamefont {Lo},
  \citenamefont {Curty},\ and\ \citenamefont {Qi}}]{lo2012measurement}%
  \BibitemOpen
  \bibfield  {author} {\bibinfo {author} {\bibfnamefont {Hoi-Kwong}\
  \bibnamefont {Lo}}, \bibinfo {author} {\bibfnamefont {Marcos}\ \bibnamefont
  {Curty}}, \ and\ \bibinfo {author} {\bibfnamefont {Bing}\ \bibnamefont
  {Qi}},\ }\bibfield  {title} {\enquote {\bibinfo {title}
  {Measurement-device-independent quantum key distribution},}\ }\href@noop {}
  {\bibfield  {journal} {\bibinfo  {journal} {Phys. Rev. Lett.}\ }\textbf
  {\bibinfo {volume} {108}},\ \bibinfo {pages} {130503} (\bibinfo {year}
  {2012})}\BibitemShut {NoStop}%
\bibitem [{\citenamefont {Braunstein}\ and\ \citenamefont
  {Pirandola}(2012)}]{BraunsteinMDI2012}%
  \BibitemOpen
  \bibfield  {author} {\bibinfo {author} {\bibfnamefont {Samuel~L.}\
  \bibnamefont {Braunstein}}\ and\ \bibinfo {author} {\bibfnamefont {Stefano}\
  \bibnamefont {Pirandola}},\ }\bibfield  {title} {\enquote {\bibinfo {title}
  {Side-channel-free quantum key distribution},}\ }\href@noop {} {\bibfield
  {journal} {\bibinfo  {journal} {Phys. Rev. Lett.}\ }\textbf {\bibinfo
  {volume} {108}},\ \bibinfo {pages} {130502} (\bibinfo {year}
  {2012})}\BibitemShut {NoStop}%
\bibitem [{\citenamefont {Wang}(2013)}]{WangThreeMDI2013}%
  \BibitemOpen
  \bibfield  {author} {\bibinfo {author} {\bibfnamefont {Xiang-Bin}\
  \bibnamefont {Wang}},\ }\bibfield  {title} {\enquote {\bibinfo {title}
  {Three-intensity decoy-state method for device-independent quantum key
  distribution with basis-dependent errors},}\ }\href@noop {} {\bibfield
  {journal} {\bibinfo  {journal} {Phys. Rev. A}\ }\textbf {\bibinfo {volume}
  {87}},\ \bibinfo {pages} {012320} (\bibinfo {year} {2013})}\BibitemShut
  {NoStop}%
\bibitem [{\citenamefont {Zhou}\ \emph {et~al.}(2016)\citenamefont {Zhou},
  \citenamefont {Yu},\ and\ \citenamefont {Wang}}]{zhouFourMDI2016}%
  \BibitemOpen
  \bibfield  {author} {\bibinfo {author} {\bibfnamefont {Yi-Heng}\ \bibnamefont
  {Zhou}}, \bibinfo {author} {\bibfnamefont {Zong-Wen}\ \bibnamefont {Yu}}, \
  and\ \bibinfo {author} {\bibfnamefont {Xiang-Bin}\ \bibnamefont {Wang}},\
  }\bibfield  {title} {\enquote {\bibinfo {title} {Making the decoy-state
  measurement-device-independent quantum key distribution practically
  useful},}\ }\href@noop {} {\bibfield  {journal} {\bibinfo  {journal} {Phys.
  Rev. A}\ }\textbf {\bibinfo {volume} {93}},\ \bibinfo {pages} {042324}
  (\bibinfo {year} {2016})}\BibitemShut {NoStop}%
\bibitem [{\citenamefont {Yin}\ \emph {et~al.}(2016{\natexlab{b}})\citenamefont
  {Yin}, \citenamefont {Chen}, \citenamefont {Yu}, \citenamefont {Liu},
  \citenamefont {You}, \citenamefont {Zhou}, \citenamefont {Chen},
  \citenamefont {Mao}, \citenamefont {Huang}, \citenamefont {Zhang},
  \citenamefont {Chen}, \citenamefont {Li}, \citenamefont {Nolan},
  \citenamefont {Zhou}, \citenamefont {Jiang}, \citenamefont {Wang},
  \citenamefont {Zhang}, \citenamefont {Wang},\ and\ \citenamefont
  {Pan}}]{MDI404}%
  \BibitemOpen
  \bibfield  {author} {\bibinfo {author} {\bibfnamefont {Hua-Lei}\ \bibnamefont
  {Yin}}, \bibinfo {author} {\bibfnamefont {Teng-Yun}\ \bibnamefont {Chen}},
  \bibinfo {author} {\bibfnamefont {Zong-Wen}\ \bibnamefont {Yu}}, \bibinfo
  {author} {\bibfnamefont {Hui}\ \bibnamefont {Liu}}, \bibinfo {author}
  {\bibfnamefont {Li-Xing}\ \bibnamefont {You}}, \bibinfo {author}
  {\bibfnamefont {Yi-Heng}\ \bibnamefont {Zhou}}, \bibinfo {author}
  {\bibfnamefont {Si-Jing}\ \bibnamefont {Chen}}, \bibinfo {author}
  {\bibfnamefont {Yingqiu}\ \bibnamefont {Mao}}, \bibinfo {author}
  {\bibfnamefont {Ming-Qi}\ \bibnamefont {Huang}}, \bibinfo {author}
  {\bibfnamefont {Wei-Jun}\ \bibnamefont {Zhang}}, \bibinfo {author}
  {\bibfnamefont {Hao}\ \bibnamefont {Chen}}, \bibinfo {author} {\bibfnamefont
  {Ming~Jun}\ \bibnamefont {Li}}, \bibinfo {author} {\bibfnamefont {Daniel}\
  \bibnamefont {Nolan}}, \bibinfo {author} {\bibfnamefont {Fei}\ \bibnamefont
  {Zhou}}, \bibinfo {author} {\bibfnamefont {Xiao}\ \bibnamefont {Jiang}},
  \bibinfo {author} {\bibfnamefont {Zhen}\ \bibnamefont {Wang}}, \bibinfo
  {author} {\bibfnamefont {Qiang}\ \bibnamefont {Zhang}}, \bibinfo {author}
  {\bibfnamefont {Xiang-Bin}\ \bibnamefont {Wang}}, \ and\ \bibinfo {author}
  {\bibfnamefont {Jian-Wei}\ \bibnamefont {Pan}},\ }\bibfield  {title}
  {\enquote {\bibinfo {title} {Measurement-device-independent quantum key
  distribution over a 404 km optical fiber},}\ }\href@noop {} {\bibfield
  {journal} {\bibinfo  {journal} {Phys. Rev. Lett.}\ }\textbf {\bibinfo
  {volume} {117}},\ \bibinfo {pages} {190501} (\bibinfo {year}
  {2016}{\natexlab{b}})}\BibitemShut {NoStop}%
\bibitem [{\citenamefont {Wang}\ \emph {et~al.}(2018)\citenamefont {Wang},
  \citenamefont {Yu},\ and\ \citenamefont {Hu}}]{wang2018twin}%
  \BibitemOpen
  \bibfield  {author} {\bibinfo {author} {\bibfnamefont {Xiang-Bin}\
  \bibnamefont {Wang}}, \bibinfo {author} {\bibfnamefont {Zong-Wen}\
  \bibnamefont {Yu}}, \ and\ \bibinfo {author} {\bibfnamefont {Xiao-Long}\
  \bibnamefont {Hu}},\ }\bibfield  {title} {\enquote {\bibinfo {title}
  {Twin-field quantum key distribution with large misalignment error},}\
  }\href@noop {} {\bibfield  {journal} {\bibinfo  {journal} {Phys. Rev. A}\
  }\textbf {\bibinfo {volume} {98}},\ \bibinfo {pages} {062323} (\bibinfo
  {year} {2018})}\BibitemShut {NoStop}%
\bibitem [{\citenamefont {Lucamarini}\ \emph {et~al.}({2018})\citenamefont
  {Lucamarini}, \citenamefont {Yuan}, \citenamefont {Dynes},\ and\
  \citenamefont {Shields}}]{LuTF2018}%
  \BibitemOpen
  \bibfield  {author} {\bibinfo {author} {\bibfnamefont {M.}~\bibnamefont
  {Lucamarini}}, \bibinfo {author} {\bibfnamefont {Z.~L.}\ \bibnamefont
  {Yuan}}, \bibinfo {author} {\bibfnamefont {J.~F.}\ \bibnamefont {Dynes}}, \
  and\ \bibinfo {author} {\bibfnamefont {A.~J.}\ \bibnamefont {Shields}},\
  }\bibfield  {title} {\enquote {\bibinfo {title} {Overcoming the rate-distance
  limit of quantum key distribution without quantum repeaters},}\ }\href@noop
  {} {\bibfield  {journal} {\bibinfo  {journal} {Nature}\ }\textbf {\bibinfo
  {volume} {{557}}},\ \bibinfo {pages} {{400--403}} (\bibinfo {year}
  {{2018}})}\BibitemShut {NoStop}%
\bibitem [{\citenamefont {Zhang}\ \emph {et~al.}(2021)\citenamefont {Zhang},
  \citenamefont {Zhou}, \citenamefont {Zhang}, \citenamefont {Li},\ and\
  \citenamefont {Wang}}]{WangTFQDS2021}%
  \BibitemOpen
  \bibfield  {author} {\bibinfo {author} {\bibfnamefont {Chun-Hui}\
  \bibnamefont {Zhang}}, \bibinfo {author} {\bibfnamefont {Xingyu}\
  \bibnamefont {Zhou}}, \bibinfo {author} {\bibfnamefont {Chun-Mei}\
  \bibnamefont {Zhang}}, \bibinfo {author} {\bibfnamefont {Jian}\ \bibnamefont
  {Li}}, \ and\ \bibinfo {author} {\bibfnamefont {Qin}\ \bibnamefont {Wang}},\
  }\bibfield  {title} {\enquote {\bibinfo {title} {Twin-field quantum digital
  signatures},}\ }\href@noop {} {\bibfield  {journal} {\bibinfo  {journal}
  {Optics Letters}\ }\textbf {\bibinfo {volume} {46}},\ \bibinfo {pages}
  {3757--3760} (\bibinfo {year} {2021})}\BibitemShut {NoStop}%
\bibitem [{\citenamefont {Gottesman}\ and\ \citenamefont
  {Lo}(2003)}]{GottesmanLo2002TWCCQKD}%
  \BibitemOpen
  \bibfield  {author} {\bibinfo {author} {\bibfnamefont {Daniel}\ \bibnamefont
  {Gottesman}}\ and\ \bibinfo {author} {\bibfnamefont {Hoi-Kwong}\ \bibnamefont
  {Lo}},\ }\bibfield  {title} {\enquote {\bibinfo {title} {Proof of security of
  quantum key distribution with two-way classical communications},}\
  }\href@noop {} {\bibfield  {journal} {\bibinfo  {journal} {IEEE Transactions
  on Information Theory}\ }\textbf {\bibinfo {volume} {49}},\ \bibinfo {pages}
  {457--475} (\bibinfo {year} {2003})}\BibitemShut {NoStop}%
\bibitem [{\citenamefont {Xu}\ \emph {et~al.}(2020)\citenamefont {Xu},
  \citenamefont {Yu}, \citenamefont {Jiang}, \citenamefont {Hu},\ and\
  \citenamefont {Wang}}]{xu2020sending}%
  \BibitemOpen
  \bibfield  {author} {\bibinfo {author} {\bibfnamefont {Hai}\ \bibnamefont
  {Xu}}, \bibinfo {author} {\bibfnamefont {Zong-Wen}\ \bibnamefont {Yu}},
  \bibinfo {author} {\bibfnamefont {Cong}\ \bibnamefont {Jiang}}, \bibinfo
  {author} {\bibfnamefont {Xiao-Long}\ \bibnamefont {Hu}}, \ and\ \bibinfo
  {author} {\bibfnamefont {Xiang-Bin}\ \bibnamefont {Wang}},\ }\bibfield
  {title} {\enquote {\bibinfo {title} {Sending-or-not-sending twin-field
  quantum key distribution: Breaking the direct transmission key rate},}\
  }\href@noop {} {\bibfield  {journal} {\bibinfo  {journal} {Phys. Rev. A}\
  }\textbf {\bibinfo {volume} {101}},\ \bibinfo {pages} {042330} (\bibinfo
  {year} {2020})}\BibitemShut {NoStop}%
\bibitem [{\citenamefont {Liu}\ \emph {et~al.}(2019)\citenamefont {Liu},
  \citenamefont {Yu}, \citenamefont {Zhang}, \citenamefont {Guan},
  \citenamefont {Chen}, \citenamefont {Zhang}, \citenamefont {Hu},
  \citenamefont {Li}, \citenamefont {Jiang}, \citenamefont {Lin}, \citenamefont
  {Chen}, \citenamefont {You}, \citenamefont {Wang}, \citenamefont {Wang},
  \citenamefont {Zhang},\ and\ \citenamefont {Pan}}]{SNS_1realsetup}%
  \BibitemOpen
  \bibfield  {author} {\bibinfo {author} {\bibfnamefont {Yang}\ \bibnamefont
  {Liu}}, \bibinfo {author} {\bibfnamefont {Zong-Wen}\ \bibnamefont {Yu}},
  \bibinfo {author} {\bibfnamefont {Weijun}\ \bibnamefont {Zhang}}, \bibinfo
  {author} {\bibfnamefont {Jian-Yu}\ \bibnamefont {Guan}}, \bibinfo {author}
  {\bibfnamefont {Jiu-Peng}\ \bibnamefont {Chen}}, \bibinfo {author}
  {\bibfnamefont {Chi}\ \bibnamefont {Zhang}}, \bibinfo {author} {\bibfnamefont
  {Xiao-Long}\ \bibnamefont {Hu}}, \bibinfo {author} {\bibfnamefont {Hao}\
  \bibnamefont {Li}}, \bibinfo {author} {\bibfnamefont {Cong}\ \bibnamefont
  {Jiang}}, \bibinfo {author} {\bibfnamefont {Jin}\ \bibnamefont {Lin}},
  \bibinfo {author} {\bibfnamefont {Teng-Yun}\ \bibnamefont {Chen}}, \bibinfo
  {author} {\bibfnamefont {Lixing}\ \bibnamefont {You}}, \bibinfo {author}
  {\bibfnamefont {Zhen}\ \bibnamefont {Wang}}, \bibinfo {author} {\bibfnamefont
  {Xiang-Bin}\ \bibnamefont {Wang}}, \bibinfo {author} {\bibfnamefont {Qiang}\
  \bibnamefont {Zhang}}, \ and\ \bibinfo {author} {\bibfnamefont {Jian-Wei}\
  \bibnamefont {Pan}},\ }\bibfield  {title} {\enquote {\bibinfo {title}
  {Experimental twin-field quantum key distribution through sending or not
  sending},}\ }\href@noop {} {\bibfield  {journal} {\bibinfo  {journal} {Phys.
  Rev. Lett.}\ }\textbf {\bibinfo {volume} {123}},\ \bibinfo {pages} {100505}
  (\bibinfo {year} {2019})}\BibitemShut {NoStop}%
\bibitem [{\citenamefont {Chen}\ \emph {et~al.}(2020)\citenamefont {Chen},
  \citenamefont {Zhang}, \citenamefont {Liu}, \citenamefont {Jiang},
  \citenamefont {Zhang}, \citenamefont {Hu}, \citenamefont {Guan},
  \citenamefont {Yu}, \citenamefont {Xu}, \citenamefont {Lin}, \citenamefont
  {Li}, \citenamefont {Chen}, \citenamefont {Li}, \citenamefont {You},
  \citenamefont {Wang}, \citenamefont {Wang}, \citenamefont {Zhang},\ and\
  \citenamefont {Pan}}]{SNS509}%
  \BibitemOpen
  \bibfield  {author} {\bibinfo {author} {\bibfnamefont {Jiu-Peng}\
  \bibnamefont {Chen}}, \bibinfo {author} {\bibfnamefont {Chi}\ \bibnamefont
  {Zhang}}, \bibinfo {author} {\bibfnamefont {Yang}\ \bibnamefont {Liu}},
  \bibinfo {author} {\bibfnamefont {Cong}\ \bibnamefont {Jiang}}, \bibinfo
  {author} {\bibfnamefont {Weijun}\ \bibnamefont {Zhang}}, \bibinfo {author}
  {\bibfnamefont {Xiao-Long}\ \bibnamefont {Hu}}, \bibinfo {author}
  {\bibfnamefont {Jian-Yu}\ \bibnamefont {Guan}}, \bibinfo {author}
  {\bibfnamefont {Zong-Wen}\ \bibnamefont {Yu}}, \bibinfo {author}
  {\bibfnamefont {Hai}\ \bibnamefont {Xu}}, \bibinfo {author} {\bibfnamefont
  {Jin}\ \bibnamefont {Lin}}, \bibinfo {author} {\bibfnamefont {Ming-Jun}\
  \bibnamefont {Li}}, \bibinfo {author} {\bibfnamefont {Hao}\ \bibnamefont
  {Chen}}, \bibinfo {author} {\bibfnamefont {Hao}\ \bibnamefont {Li}}, \bibinfo
  {author} {\bibfnamefont {Lixing}\ \bibnamefont {You}}, \bibinfo {author}
  {\bibfnamefont {Zhen}\ \bibnamefont {Wang}}, \bibinfo {author} {\bibfnamefont
  {Xiang-Bin}\ \bibnamefont {Wang}}, \bibinfo {author} {\bibfnamefont {Qiang}\
  \bibnamefont {Zhang}}, \ and\ \bibinfo {author} {\bibfnamefont {Jian-Wei}\
  \bibnamefont {Pan}},\ }\bibfield  {title} {\enquote {\bibinfo {title}
  {Sending-or-not-sending with independent lasers: Secure twin-field quantum
  key distribution over 509 km},}\ }\href@noop {} {\bibfield  {journal}
  {\bibinfo  {journal} {Phys. Rev. Lett.}\ }\textbf {\bibinfo {volume} {124}},\
  \bibinfo {pages} {070501} (\bibinfo {year} {2020})}\BibitemShut {NoStop}%
\bibitem [{\citenamefont {Pittaluga}\ \emph {et~al.}({2021})\citenamefont
  {Pittaluga}, \citenamefont {Minder}, \citenamefont {Lucamarini},
  \citenamefont {Sanzaro}, \citenamefont {Woodward}, \citenamefont {Li},
  \citenamefont {Yuan},\ and\ \citenamefont {Shields}}]{SNS600}%
  \BibitemOpen
  \bibfield  {author} {\bibinfo {author} {\bibfnamefont {Mirko}\ \bibnamefont
  {Pittaluga}}, \bibinfo {author} {\bibfnamefont {Mariella}\ \bibnamefont
  {Minder}}, \bibinfo {author} {\bibfnamefont {Marco}\ \bibnamefont
  {Lucamarini}}, \bibinfo {author} {\bibfnamefont {Mirko}\ \bibnamefont
  {Sanzaro}}, \bibinfo {author} {\bibfnamefont {Robert}\ \bibnamefont
  {Woodward}, \bibfnamefont {I}}, \bibinfo {author} {\bibfnamefont {Ming-Jun}\
  \bibnamefont {Li}}, \bibinfo {author} {\bibfnamefont {Zhiliang}\ \bibnamefont
  {Yuan}}, \ and\ \bibinfo {author} {\bibfnamefont {Andrew~J.}\ \bibnamefont
  {Shields}},\ }\bibfield  {title} {\enquote {\bibinfo {title} {{600-km
  repeater-like quantum communications with dual-band stabilization}},}\
  }\href@noop {} {\bibfield  {journal} {\bibinfo  {journal} {Nature Photonics}\
  }\textbf {\bibinfo {volume} {{15}}},\ \bibinfo {pages} {{530+}} (\bibinfo
  {year} {{2021}})}\BibitemShut {NoStop}%
\bibitem [{\citenamefont {Liu}\ \emph {et~al.}(2021)\citenamefont {Liu},
  \citenamefont {Jiang}, \citenamefont {Zhu}, \citenamefont {Zou},
  \citenamefont {Yu}, \citenamefont {Hu}, \citenamefont {Xu}, \citenamefont
  {Ma}, \citenamefont {Han}, \citenamefont {Chen}, \citenamefont {Dai},
  \citenamefont {Tang}, \citenamefont {Zhang}, \citenamefont {Li},
  \citenamefont {You}, \citenamefont {Wang}, \citenamefont {Hua}, \citenamefont
  {Hu}, \citenamefont {Zhang}, \citenamefont {Zhou}, \citenamefont {Zhang},
  \citenamefont {Wang}, \citenamefont {Chen},\ and\ \citenamefont
  {Pan}}]{SNS428}%
  \BibitemOpen
  \bibfield  {author} {\bibinfo {author} {\bibfnamefont {Hui}\ \bibnamefont
  {Liu}}, \bibinfo {author} {\bibfnamefont {Cong}\ \bibnamefont {Jiang}},
  \bibinfo {author} {\bibfnamefont {Hao-Tao}\ \bibnamefont {Zhu}}, \bibinfo
  {author} {\bibfnamefont {Mi}~\bibnamefont {Zou}}, \bibinfo {author}
  {\bibfnamefont {Zong-Wen}\ \bibnamefont {Yu}}, \bibinfo {author}
  {\bibfnamefont {Xiao-Long}\ \bibnamefont {Hu}}, \bibinfo {author}
  {\bibfnamefont {Hai}\ \bibnamefont {Xu}}, \bibinfo {author} {\bibfnamefont
  {Shizhao}\ \bibnamefont {Ma}}, \bibinfo {author} {\bibfnamefont {Zhiyong}\
  \bibnamefont {Han}}, \bibinfo {author} {\bibfnamefont {Jiu-Peng}\
  \bibnamefont {Chen}}, \bibinfo {author} {\bibfnamefont {Yunqi}\ \bibnamefont
  {Dai}}, \bibinfo {author} {\bibfnamefont {Shi-Biao}\ \bibnamefont {Tang}},
  \bibinfo {author} {\bibfnamefont {Weijun}\ \bibnamefont {Zhang}}, \bibinfo
  {author} {\bibfnamefont {Hao}\ \bibnamefont {Li}}, \bibinfo {author}
  {\bibfnamefont {Lixing}\ \bibnamefont {You}}, \bibinfo {author}
  {\bibfnamefont {Zhen}\ \bibnamefont {Wang}}, \bibinfo {author} {\bibfnamefont
  {Yong}\ \bibnamefont {Hua}}, \bibinfo {author} {\bibfnamefont {Hongkun}\
  \bibnamefont {Hu}}, \bibinfo {author} {\bibfnamefont {Hongbo}\ \bibnamefont
  {Zhang}}, \bibinfo {author} {\bibfnamefont {Fei}\ \bibnamefont {Zhou}},
  \bibinfo {author} {\bibfnamefont {Qiang}\ \bibnamefont {Zhang}}, \bibinfo
  {author} {\bibfnamefont {Xiang-Bin}\ \bibnamefont {Wang}}, \bibinfo {author}
  {\bibfnamefont {Teng-Yun}\ \bibnamefont {Chen}}, \ and\ \bibinfo {author}
  {\bibfnamefont {Jian-Wei}\ \bibnamefont {Pan}},\ }\bibfield  {title}
  {\enquote {\bibinfo {title} {Field test of twin-field quantum key
  distribution through sending-or-not-sending over 428 km},}\ }\href@noop {}
  {\bibfield  {journal} {\bibinfo  {journal} {Phys. Rev. Lett.}\ }\textbf
  {\bibinfo {volume} {126}},\ \bibinfo {pages} {250502} (\bibinfo {year}
  {2021})}\BibitemShut {NoStop}%
\bibitem [{\citenamefont {Chen}\ \emph {et~al.}({2021})\citenamefont {Chen},
  \citenamefont {Zhang}, \citenamefont {Liu}, \citenamefont {Jiang},
  \citenamefont {Zhang}, \citenamefont {Han}, \citenamefont {Ma}, \citenamefont
  {Hu}, \citenamefont {Li}, \citenamefont {Liu}, \citenamefont {Zhou},
  \citenamefont {Jiang}, \citenamefont {Chen}, \citenamefont {Li},
  \citenamefont {You}, \citenamefont {Wang}, \citenamefont {Wang},
  \citenamefont {Zhang},\ and\ \citenamefont {Pan}}]{SNS511}%
  \BibitemOpen
  \bibfield  {author} {\bibinfo {author} {\bibfnamefont {Jiu-Peng}\
  \bibnamefont {Chen}}, \bibinfo {author} {\bibfnamefont {Chi}\ \bibnamefont
  {Zhang}}, \bibinfo {author} {\bibfnamefont {Yang}\ \bibnamefont {Liu}},
  \bibinfo {author} {\bibfnamefont {Cong}\ \bibnamefont {Jiang}}, \bibinfo
  {author} {\bibfnamefont {Wei-Jun}\ \bibnamefont {Zhang}}, \bibinfo {author}
  {\bibfnamefont {Zhi-Yong}\ \bibnamefont {Han}}, \bibinfo {author}
  {\bibfnamefont {Shi-Zhao}\ \bibnamefont {Ma}}, \bibinfo {author}
  {\bibfnamefont {Xiao-Long}\ \bibnamefont {Hu}}, \bibinfo {author}
  {\bibfnamefont {Yu-Huai}\ \bibnamefont {Li}}, \bibinfo {author}
  {\bibfnamefont {Hui}\ \bibnamefont {Liu}}, \bibinfo {author} {\bibfnamefont
  {Fei}\ \bibnamefont {Zhou}}, \bibinfo {author} {\bibfnamefont {Hai-Feng}\
  \bibnamefont {Jiang}}, \bibinfo {author} {\bibfnamefont {Teng-Yun}\
  \bibnamefont {Chen}}, \bibinfo {author} {\bibfnamefont {Hao}\ \bibnamefont
  {Li}}, \bibinfo {author} {\bibfnamefont {Li-Xing}\ \bibnamefont {You}},
  \bibinfo {author} {\bibfnamefont {Zhen}\ \bibnamefont {Wang}}, \bibinfo
  {author} {\bibfnamefont {Xiang-Bin}\ \bibnamefont {Wang}}, \bibinfo {author}
  {\bibfnamefont {Qiang}\ \bibnamefont {Zhang}}, \ and\ \bibinfo {author}
  {\bibfnamefont {Jian-Wei}\ \bibnamefont {Pan}},\ }\bibfield  {title}
  {\enquote {\bibinfo {title} {{Twin-field quantum key distribution over a
  511km optical fibre linking two distant metropolitan areas}},}\ }\href@noop
  {} {\bibfield  {journal} {\bibinfo  {journal} {Nature Photonics}\ }\textbf
  {\bibinfo {volume} {{15}}},\ \bibinfo {pages} {{570+}} (\bibinfo {year}
  {{2021}})}\BibitemShut {NoStop}%
\bibitem [{\citenamefont {Wang}\ \emph {et~al.}(2019)\citenamefont {Wang},
  \citenamefont {Hu},\ and\ \citenamefont {Yu}}]{wang2019practical}%
  \BibitemOpen
  \bibfield  {author} {\bibinfo {author} {\bibfnamefont {Xiang-Bin}\
  \bibnamefont {Wang}}, \bibinfo {author} {\bibfnamefont {Xiao-Long}\
  \bibnamefont {Hu}}, \ and\ \bibinfo {author} {\bibfnamefont {Zong-Wen}\
  \bibnamefont {Yu}},\ }\bibfield  {title} {\enquote {\bibinfo {title}
  {Practical long-distance side-channel-free quantum key distribution},}\
  }\href@noop {} {\bibfield  {journal} {\bibinfo  {journal} {Phys. Rev.
  Applied}\ }\textbf {\bibinfo {volume} {12}},\ \bibinfo {pages} {054034}
  (\bibinfo {year} {2019})}\BibitemShut {NoStop}%
\bibitem [{\citenamefont {Wang}\ \emph {et~al.}(2015)\citenamefont {Wang},
  \citenamefont {Cai}, \citenamefont {Ren},\ and\ \citenamefont
  {Zhang}}]{wang2015longmessages}%
  \BibitemOpen
  \bibfield  {author} {\bibinfo {author} {\bibfnamefont {Tian-Yin}\
  \bibnamefont {Wang}}, \bibinfo {author} {\bibfnamefont {Xiao-Qiu}\
  \bibnamefont {Cai}}, \bibinfo {author} {\bibfnamefont {Yan-Li}\ \bibnamefont
  {Ren}}, \ and\ \bibinfo {author} {\bibfnamefont {Rui-Ling}\ \bibnamefont
  {Zhang}},\ }\bibfield  {title} {\enquote {\bibinfo {title} {Security of
  quantum digital signatures for classical messages},}\ }\href@noop {}
  {\bibfield  {journal} {\bibinfo  {journal} {Scientific Reports}\ }\textbf
  {\bibinfo {volume} {5}},\ \bibinfo {pages} {9231} (\bibinfo {year}
  {2015})}\BibitemShut {NoStop}%
\bibitem [{\citenamefont {Serfling}(1974)}]{Serfling1974}%
  \BibitemOpen
  \bibfield  {author} {\bibinfo {author} {\bibfnamefont {R.~J.}\ \bibnamefont
  {Serfling}},\ }\bibfield  {title} {\enquote {\bibinfo {title} {Probability
  inequalities for the sum in sampling without replacement},}\ }\href@noop {}
  {\bibfield  {journal} {\bibinfo  {journal} {The Annals of Statistics}\
  }\textbf {\bibinfo {volume} {2}},\ \bibinfo {pages} {39--48} (\bibinfo {year}
  {1974})}\BibitemShut {NoStop}%
\bibitem [{\citenamefont {Yu}\ \emph {et~al.}({2019})\citenamefont {Yu},
  \citenamefont {Hu}, \citenamefont {Jiang}, \citenamefont {Xu},\ and\
  \citenamefont {Wang}}]{Yu2019SNSdecoy}%
  \BibitemOpen
  \bibfield  {author} {\bibinfo {author} {\bibfnamefont {Zong-Wen}\
  \bibnamefont {Yu}}, \bibinfo {author} {\bibfnamefont {Xiao-Long}\
  \bibnamefont {Hu}}, \bibinfo {author} {\bibfnamefont {Cong}\ \bibnamefont
  {Jiang}}, \bibinfo {author} {\bibfnamefont {Hai}\ \bibnamefont {Xu}}, \ and\
  \bibinfo {author} {\bibfnamefont {Xiang-Bin}\ \bibnamefont {Wang}},\
  }\bibfield  {title} {\enquote {\bibinfo {title} {Sending-or-not-sending
  twin-field quantum key distribution in practice},}\ }\href@noop {} {\bibfield
   {journal} {\bibinfo  {journal} {Scientific reports}\ }\textbf {\bibinfo
  {volume} {{9}}} (\bibinfo {year} {{2019}})}\BibitemShut {NoStop}%
\bibitem [{\citenamefont {Jiang}\ \emph {et~al.}(2019)\citenamefont {Jiang},
  \citenamefont {Yu}, \citenamefont {Hu},\ and\ \citenamefont
  {Wang}}]{Jiang2019SNSfinite}%
  \BibitemOpen
  \bibfield  {author} {\bibinfo {author} {\bibfnamefont {Cong}\ \bibnamefont
  {Jiang}}, \bibinfo {author} {\bibfnamefont {Zong-Wen}\ \bibnamefont {Yu}},
  \bibinfo {author} {\bibfnamefont {Xiao-Long}\ \bibnamefont {Hu}}, \ and\
  \bibinfo {author} {\bibfnamefont {Xiang-Bin}\ \bibnamefont {Wang}},\
  }\bibfield  {title} {\enquote {\bibinfo {title} {Unconditional security of
  sending or not sending twin-field quantum key distribution with finite
  pulses},}\ }\href@noop {} {\bibfield  {journal} {\bibinfo  {journal} {Phys.
  Rev. Applied}\ }\textbf {\bibinfo {volume} {12}},\ \bibinfo {pages} {024061}
  (\bibinfo {year} {2019})}\BibitemShut {NoStop}%
\bibitem [{\citenamefont {Chernoff}(1952)}]{chernoff}%
  \BibitemOpen
  \bibfield  {author} {\bibinfo {author} {\bibfnamefont {Herman}\ \bibnamefont
  {Chernoff}},\ }\bibfield  {title} {\enquote {\bibinfo {title} {A measure of
  asymptotic efficiency for tests of a hypothesis based on the sum of
  observations},}\ }\href@noop {} {\bibfield  {journal} {\bibinfo  {journal}
  {The Annals of Mathematical Statistics}\ ,\ \bibinfo {pages} {493--507}}
  (\bibinfo {year} {1952})}\BibitemShut {NoStop}%
\end{thebibliography}%

\end{document}